\documentclass[journal]{IEEEtran}
\usepackage{amsfonts}
\newtheorem{remark}{Remark}
\usepackage{bm}
\usepackage{amsmath}
\usepackage{makecell}
\usepackage{amssymb}
\usepackage{array}
\usepackage{cases}
\usepackage{algorithm}
\usepackage{algorithmic}
\usepackage{enumerate}
\usepackage{color,xcolor}
\usepackage{graphicx}
\usepackage{multirow,tabularx}
\usepackage{subfigure}
\usepackage{multirow}
\usepackage{hyperref}
\usepackage[compress]{cite}
\usepackage{setspace}

\makeatletter

\newcommand{\Rmnum}[1]{\expandafter\@slowromancap\romannumeral #1@}
\usepackage[vlined, ruled, algo2e]{algorithm2e}
\makeatother
 0

\begin{document}
\title{DNN-Aided Block Sparse Bayesian Learning for User Activity Detection and Channel Estimation in Grant-Free Non-Orthogonal Random Access}

\author{
\IEEEauthorblockN{Zhaoji~Zhang, Ying~Li, \IEEEmembership{Member, IEEE}, Chongwen Huang, Qinghua Guo, \IEEEmembership{Senior Member, IEEE}, Chau Yuen, \IEEEmembership{Senior Member, IEEE}, and Yong Liang Guan, \IEEEmembership{Senior Member, IEEE}}
	\thanks{Copyright (c) 2015 IEEE. Personal use of this material is permitted. However, permission to use this material for any other purposes must be obtained from the IEEE by sending a request to pubs-permissions@ieee.org.}
	\thanks{Part of this work was presented at 2019 IEEE VTS Asia Pacific Wireless Communications Symposium \cite{APWCS}. This work is supported in part by NSFC grant 61971333, NSFC grant 61671345, NSFC grant 61750110529, and A*STAR AME IAF PP Grant-A19D6a0053. (\emph{Corresponding author: Ying Li})}
    \thanks{Z. Zhang and Y. Li are with Xidian University, Xi'an, 710071, China
	(e-mail: zjzhang\_1@stu.xidian.edu.cn; yli@mail.xidian.edu.cn).}
	\thanks{C. Huang and C. Yuen are with Singapore University of Technology and Design, Singapore 119613 (e-mail: chongwen\_huang@mymail.sutd.edu.sg; yuenchau@sutd.edu.sg).}
	\thanks{Q. Guo is with the School of Electrical, Computer and Telecommunications Engineering, University of Wollongong, Wollongong, NSW, Australia (e-mail: qguo@uow.edu.au).}
	\thanks{Y. L. Guan is with the School of Electrical and Electronic Engineering, Nanyang Technological University, Singapore 639798 (e-mail: eylguan@ntu.edu.sg).}}

\maketitle
\begin{abstract}
In the upcoming Internet-of-Things (IoT) era, the communication is often featured by massive connection, sporadic transmission, and small-sized data packets, which poses new requirements on the delay expectation and resource allocation efficiency of the Random Access (RA) mechanisms of the IoT communication stack. A grant-free non-orthogonal random access (NORA) system is considered in this paper, which could simultaneously reduce the access delay and support more Machine Type Communication (MTC) devices with limited resources. In order to address the joint user activity detection (UAD) and channel estimation (CE) problem in the grant-free NORA system, we propose a deep neural network-aided message passing-based block sparse Bayesian learning (DNN-MP-BSBL) algorithm. In the DNN-MP-BSBL algorithm, the iterative message passing process is transferred from a factor graph to a deep neural network (DNN). Weights are imposed on the messages in the DNN and trained to minimize the estimation error. It is shown that the trained weights could alleviate the convergence problem of the MP-BSBL algorithm, especially on crowded RA scenarios. Simulation results show that the proposed DNN-MP-BSBL algorithm could improve the UAD and CE accuracy with a smaller number of iterations, indicating its advantages for low-latency grant-free NORA systems.
\end{abstract}
\begin{IEEEkeywords}
deep neural network, sparse Bayesian learning, grant-free, user activity detection, channel estimation
\end{IEEEkeywords}


%

\section{Introduction}
\IEEEPARstart{P}{roviding} efficient support for the Internet of Things (IoT) and Industry IoT (IIoT) \cite{IIOT} is one of the major objectives for the fifth generation (5G) and the next-generation 6G cellular wireless communication \cite{6G}. In order to realize the physical information exchange in IoT, Machine-to-Machine (M2M) communication is anticipated to support billions of Machine Type Communication (MTC) devices \cite{2020}. In addition, for most IoT applications such as smart metering and intelligent transportation, the MTC devices are intermittently activated with a low probability, and the size of the data packets transmitted by active devices is relatively small  \cite{servicetype,servicetypetwo}. Therefore, the random access (RA) process for M2M communications in IoT is characterized by massive connection and sporadic transmission, as well as small-sized data packets.

Confronted with the characteristics described above, conventional orthogonal multiple access (OMA) schemes, where orthogonal resource blocks (RBs) are assigned to each device,   becomes infeasible due to its extremely low resource efficiency. To address this problem, different RA schemes have been proposed, where activated devices can share or contend for the uplink resources, and these RA schemes can be generally categorized into two types: grant-based RA \cite{ACB,seperation,dynamicone,morePA2,reusePA,TVTearly,TVTCARA} and grant-free RA \cite{Aloha,CRDSA,CSA,frameless,SCCSA,blockCS,CRAN,2015ICC,ICASSP,AMPmassive,AMPmassive2,SBLAMP,LSAMPSBL,DLGFNOMA,MPBSBL2019,MPBSBL}.
\subsection{Grant-Based Random Access}
In grant-based RA schemes, activated users contend for the RBs by transmitting a preamble sequence to BS. At the BS, a RB is assigned to the activated user, whose preamble sequence is received and accepted by the BS.

One problem with the grant-based RA schemes is that the RB is wasted when more than one active device transmit the same preamble sequence. This problem is worsened with the massive connection in crowded RA scenarios. Some solutions were proposed to alleviate the RA congestion by reducing the collision probability, such as the Access Class Barring (ACB) scheme \cite{ACB}, delicate splitting of the RA preamble set \cite{seperation}, and automatic configuration of the RA parameters \cite{dynamicone}. Other works addressed the preamble shortage \cite{morePA2,reusePA} or the early preamble collision detection to avoid RB wastage \cite{TVTearly, TVTCARA}.

However, the RB wastage cannot be fully avoided by grant-based RA schemes, especially in crowded RA scenarios, which results in low resource efficiency. Furthermore, a handshaking process is required between the BS and active devices to recognize the contention winner, which undermines the uplink transmission efficiency of small data packets.
\subsection{Grant-Free Random Access}
In grant-free RA schemes, activated users directly transmit data packets with encapsulated pilot sequences over shared uplink resources. Therefore, no grant is required from the BS for uplink transmission, which improves the transmission efficiency, especially for small-sized data packets. Note that in grant-based RA, the BS can acquire the ID of active users during the handshaking process, so that the activity of each user can be easily recognized. By contrast, in grant-free RA, the BS has no priori knowledge of each user's activity before the transmission of data packets. Therefore, it is necessary to exploit encapsulated pilot sequences for user activity detection (UAD) and channel estimation (CE), so that the BS can identify each active user and further accomplish data decoding. Different grant-free RA schemes are explained as follows.
\subsubsection{Slotted ALOHA based grant-free RA Schemes}
In slotted ALOHA based grant-free RA schemes, activated devices directly transmit data packets in randomly chosen slots while data packets experiencing no slot collision can be correctly decoded \cite{Aloha}. In order to improve the throughput of slotted ALOHA protocols, the contention resolution diversity slotted ALOHA (CRDSA) protocol \cite{CRDSA} was proposed, where each activated device sends two replicas of the same data packet and successive interference cancellation (SIC) technique is employed to recover the collided replicas. The CRDSA protocol introduced a (2,1) repetition code to the conventional slotted ALOHA protocol while the coded slotted ALOHA (CSA) protocol \cite{CSA} further combined the general packet erasure codes with SIC. According to the analogy between the channel codes and the data packet replica, the frameless IRSA protocol \cite{frameless} and the spatially coupled RA protocol \cite{SCCSA} were proposed, corresponding to the rateless codes and the spatially coupled LDPC codes, respectively.

Although different modifications have been proposed to improve the throughput of slotted ALOHA protocols, the severe slot collision always leads to the failure of SIC, which hinders the application of slotted ALOHA based protocols in crowded RA scenarios.
\subsubsection{Compressed sensing-based grant-free RA schemes}
Due to the sporadic transmission feature of M2M communications, the RA process can be formulated as a sparse signal recovery problem. Therefore, different compressed sensing (CS) algorithms employ pilot sequences to accomplish the user activity detection (UAD) and/or channel estimation (CE) problem. For example, a block CS algorithm \cite{blockCS} was proposed for distributed UAD and resource allocation, based on the clustering of devices. The joint UAD and CE problem was addressed by a modified Bayesian compressed sensing algorithm \cite{CRAN} for the cloud radio access network (C-RAN). In addition, the powerful approximate message passing (AMP) algorithm was employed for the joint UAD and CE problem when the BS is equipped either with a single antenna \cite{2015ICC, ICASSP} or with multiple antennas \cite{AMPmassive,AMPmassive2}.
\subsubsection{Sparse Bayesian learning-based grant-free RA schemes}
Different from CS-based RA schemes, the sparse Bayesian learning (SBL) algorithm further considers the prior hyper-parameter of the sparse signal to address the joint UAD and CE problem. The Expectation Maximization (EM) method was employed by the AMP-SBL algorithm \cite{SBLAMP} to update the sparse signal and the hyper-parameter iteratively. A least square (LS)-based AMP-SBL (LS-AMP-SBL) algorithm \cite{LSAMPSBL} was proposed to recover the sparse signal in three steps, i.e., the AMP-SBL first provides a coarse estimate, followed by the derivation for the support of the sparse channel, while the final estimate is derived by the support and the LS estimation. The joint UAD and data decoding task for a grant-free NOMA system was modeled and addressed by a deep auto-encoder in \cite{DLGFNOMA}, which exploits the prior information of user activation probability and optimizes the choice of spreading signature. However, this auto-encoder is trained based on known channel state information at receiver and a sufficiently static channel, which impedes it application to the rapidly changing environment. Recently, a message-passing receiver design was proposed for the joint channel estimation and data decoding in uplink grant-free SCMA systems \cite{MPBSBL2019}, exploiting variational Bayesian inference to extract the sparse signals. In addition, a message passing-based block sparse Bayesian learning (MP-BSBL) algorithm \cite{MPBSBL} was proposed for a grant-free NOMA system. In both \cite{MPBSBL2019} and \cite{MPBSBL}, non-zero elements in the sparse signals are updated with Gaussian message passing \cite{GMPID,GMPCW,GMPCW2} and the hyper-parameters are updated with mean-field (MF) message passing \cite{MF}.

The Gaussian message passing \cite{GMPID,GMPCW,GMPCW2} is constructed on a factor graph, and it is proven to be optimal on cycle-free graphs. However, this factor graph is densely-connected under crowded RA scenarios, which causes correlated messages and convergence problem for the Gaussian message passing in MP-BSBL algorithm \cite{MPBSBL}. As a result, the CE and UAD accuracy of the MP-BSBL algorithm is undermined under crowded RA scenarios. To address this problem, we impose weighting parameters on the Gaussian messages and MF message update in the MP-BSBL algorithm. In order to facilitate the training of these weighting parameters, the edge-type message passing process on the factor graph is transformed to a node-type one in the deep neural network (DNN) \cite{DNNBCH,AccessRA}. Then these weights in the DNN can be trained in a data-driven manner to mitigate the message correlation and the convergence problem in the MP-BSBL algorithm.  In this way, higher UAD and CE accuracy can be achieved for crowded RA scenarios.
\begin{figure*}
\centering
\includegraphics[width=1\linewidth]{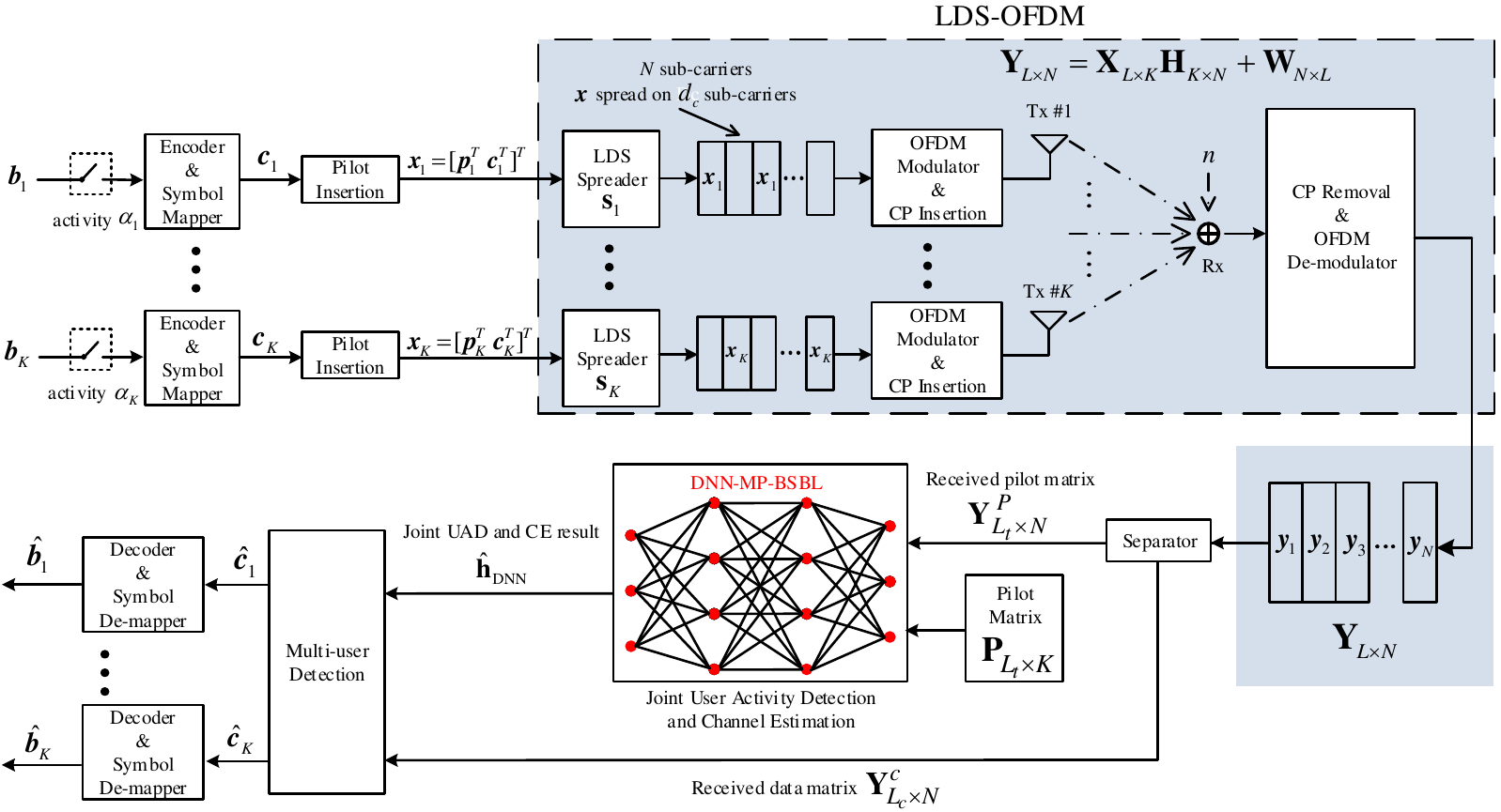}
\caption{LDS-OFDM based grant-free NORA system model. The joint UAD and CE problem is sloved by the proposed DNN-MP-BSBL algorithm, which processes the first $L_t$ received signals of the pilot sequences, while the MUD module is employed to process the remaining $L_c$ received signals to detect the data transmitted from active users.}
\label{SystemModel}
\end{figure*}
\begin{table}
\renewcommand\arraystretch{1.0}
\caption{Frequently used abbreviations and corresponding meaning.}
\centering
\begin{tabular}{c|c}
  \Xhline{1.2pt}
Abbr.&Meaning\\\hline
DNN&Deep Neural Network\\\hline
RA&Random Access\\\hline
NORA&Non-Orthogonal Random Access\\\hline
UAD&User Activity Detection\\\hline
CE&Channel Estimation\\\hline
\multirow{2}{*}{MP-BSBL \cite{MPBSBL}}&Message Passing-Based Block\\&Sparse Bayesian Learning.\\\hline
DNN-MP-BSBL&DNN-Aided MP-BSBL\\\hline
MF&Mean Field\\\hline
LDS&Low-Density Signature\\\hline
NMSE&Normalized Mean Square Error\\\hline
  \Xhline{1.2pt}
\end{tabular}
\label{abbr}
\end{table}
\subsection{Contributions}
In this paper, we focus on an orthogonal frequency division multiplexing system with low-density signature (LDS-OFDM) \cite{LDSOFDM,LDSOFDM2}, where devices perform grant-free RA once they are activated. A deep neural network-aided message passing-based block sparse Bayesian learning (DNN-MP-BSBL) algorithm is proposed in this paper to perform joint UAD and CE for the grant-free RA procedure. The iterative message passing process of the MP-BSBL algorithm \cite{MPBSBL} is transferred from a factor graph to a neural network. Weights are imposed on the messages passing in the neural network and further trained to minimize the estimation error. It is shown that the trained weights could alleviate the convergence problem of the MP-BSBL algorithm, especially in crowded RA scenarios. As a result, the DNN-MP-BSBL algorithm could improve the UAD and CE accuracy with a smaller number of iterations. In addition, the training of the DNN is conducted offline, and negligible online computational complexity is additionally sacrificed for the performance improvement by the DNN. This guarantees the feasibility of the proposed DNN-MP-BSBL algorithm for crowded NORA systems with real-time implementation and low-latency requirement.

The rest of this paper is organized as follows. The system model and the MP-BSBL algorithm are presented in Section \ref{ModelProblem}. The DNN structure for the DNN-MP-BSBL algorithm is illustrated in Section \ref{DNN_SBSL}, where the weighted message passing is explained in details. Simulation results are given in Section \ref{simulationSec} to verify the UAD and CE accuracy of the proposed DNN-MP-BSBL algorithm. Finally, Section \ref{conclusions} concludes this paper.  For reading convenience, the frequently-used abbreviations in this paper are summarized in Table. \ref{abbr}.
\section{Joint UAD and CE by MP-BSBL}\label{ModelProblem}
\subsection{System Model}
As shown in Fig. \ref{SystemModel}, a LDS-OFDM communication system is considered. It is assumed that there are $N$ sub-carriers and $K$ users, while each user is activated with probability $P_a$. For each active user $k$, its binary information sequence $\mathbf{b}_k$ is encoded and further mapped into a QAM symbol sequence $\mathbf{c}_k\in\mathbb{C}^{L_c\times1}$ with length $L_c$. One unique pilot sequence $\mathbf{p}_k$ with length $L_t$ is allocated for user $k$, and $\mathbf{p}_k$ is further inserted into the transmitted symbol sequence $\mathbf{x}_k$, i.e., $\mathbf{x}_k=[\mathbf{p}_k^T \ \mathbf{c}_k^T]^T$. Therefore, the length $L$ of $\mathbf{x}_k$ is $L=L_t+L_c$.  Then each symbol in $\mathbf{x}_k$ is spread onto $d_c$ sub-carriers, using a low-density spreading pattern $\mathbf{s}_k$, which is a sparse vector with length $N$ and $d_c$ non-zero elements. In this way, the same sequence $\mathbf{x}_k$ is transmitted on $d_c$ different sub-carriers.

The LDS spreaders for all the $K$ users are represented by an LDS spreading matrix $\mathbf{S}=[\mathbf{s}_1,...,\mathbf{s}_K]$, which is assumed known at the receiver. We consider a regular LDS spreading matrix, i.e., the column degree $d_c$ and the row degree $d_r$ in $\mathbf{S}$ are constant. According to the regular structure of $\mathbf{S}$, each sub-carrier is shared by $d_r$ potential users, with $d_{r}=(K/N)d_{c}$. When multiple users are activated on the same sub-carrier, the grant-free RA procedure is conducted in a non-orthogonal multiple access (NOMA) manner. Therefore, the system model in Fig. \ref{SystemModel} is termed a LDS-OFDM based grant-free non-orthogonal random access (NORA) system.

Before transmission, the cyclic prefix (CP) is inserted after the OFDM modulator to avoid the inter-symbol interference (ISI). At the receiver, the CP is removed, and after the OFDM demodulator, we obtain the frequency-domain received signal matrix $\mathbf{Y}_{L\times N}=[\mathbf{y}_1,\ldots,\mathbf{y}_n,\ldots,\mathbf{y}_N]$, where $\mathbf{y}_n$ is the received signal vector representing the $L$ received symbols on the $n$-th sub-carrier.

Now we discuss the channel between the input and the output of this LDS-OFDM system. The transmitted sequences of all the $K$ users are organized in an input matrix $\mathbf{X}_{L\times K}=[\mathbf{x}_1,\ldots,\mathbf{x}_k,\ldots,\mathbf{x}_K]$, while the user activity is equivalently incorporated in the channel matrix $\mathbf{H}_{K\times N}$. Specifically, we consider the \emph{effective} channel gain vector $\mathbf{h}_k$ of user $k$ to integrate the \emph{actual} channel gain vector $\mathbf{g}_k$ with the sparse spreading pattern $\mathbf{s}_k$. That is,
\begin{equation}\label{effvhannel}
\begin{split}
\mathbf{h}_k&=\left[h_{1k},\ldots,h_{nk},\ldots,h_{Nk}\right]^T\\
&=\left[g_{1k}s_{1k},\ldots,g_{nk}s_{nk},\ldots,g_{Nk}s_{Nk}\right]^T,
\end{split}
\end{equation}
where the $n$-th element $g_{nk}$ of $\mathbf{g}_k$ represents the actual channel gain of user $k$ on the $n$-th sub-carrier, and the $n$-th element $s_{nk}$ of $\mathbf{s}_k$ is non-zero if and only if user $k$ chooses to conduct transmission on the $n$-th sub-carrier. Therefore, the effective channel gain vector $\mathbf{h}_k$ is a sparse vector with $d_c$ non-zero elements. Furthermore, we incorporate user activity in the channel matrix, i.e., $\mathbf{H}_{K\times N}=[\alpha_1\mathbf{h}_1,...,\alpha_K\mathbf{h}_K]^T$, where the activity indicator $\alpha_k=0$ if user $k$ is inactive. Otherwise, $\alpha_k=1$. In this way, we can obtain
\begin{equation}\label{received}
\mathbf{Y}_{L\times N}=\mathbf{X}_{L\times K}\mathbf{H}_{K\times N}+\mathbf{W}_{N\times L},
\end{equation}
where $L=L_t+L_c$, the $(l,n)$-th entry of $\mathbf{Y}$ represents the $l$-th received symbol on the $n$-th sub-carrier, the $(l,k)$-th entry of $\mathbf{X}$ represents the $l$-th transmitted symbol of the $k$-th user. The entries in the additive white Gaussian noise (AWGN) matrix $\mathbf{W}$ are assumed i.i.d with noise variance $\sigma_{w}^2$.

According to the composition of $\mathbf{x}_k$, $\mathbf{Y}$ can be decomposed as $\mathbf{Y}_{L\times N}=\left[(\mathbf{Y}^P_{L_t\times N})^T\ (\mathbf{Y}^c_{L_c\times N})^T \right]^T$, where $\mathbf{Y}^P$ and $\mathbf{Y}^c$ represent the received signal matrices with respect to (w.r.t.) the pilot sequences and the data sequences, respectively. After the separator, $\mathbf{Y}^P$ is further processed by the proposed DNN-MP-BSBL algorithm for the joint UAD and CE. According to the output of the DNN-MP-BSBL algorithm $\hat{\mathbf{h}}_{\text{DNN}}$, $\mathbf{Y}^c$ is fed into the multi-user detection (MUD) module to detect and decode the information bits of active users.
\subsection{Problem Formulation}
In this paper, we mainly consider the received pilot matrix $\mathbf{Y}^P$ and solve the joint UAD and CE problem. Analogous to (\ref{received}), $\mathbf{Y}^P$ can be formulated as
\begin{equation}\label{Pilotreceived}
\mathbf{Y}^P_{L_t\times N}=\mathbf{P}_{L_t\times K}\mathbf{H}_{K\times N}+\mathbf{W}_{N\times L_t},
\end{equation}
where the $(l,n)$-th entry of $\mathbf{Y}^P$ represents the $l$-th received signal on the $n$-th sub-carrier, and the $(l,k)$-th entry of $\mathbf{P}$ represents the $l$-th pilot symbol of the $k$-th user. The pilot matrix $\mathbf{P}$ is assumed known to the receiver. Then we perform vectorization on the transpose of $\mathbf{Y}^P$ as in \cite{MPBSBL},
\begin{equation}\label{Vectorization}
\begin{split}
\mathbf{y}&=vec([{\mathbf{Y}^P}]^T)=(\mathbf{P}\otimes\mathbf{I}_N)vec(\mathbf{H}^T)+\mathbf{w}\\
&=\!\!\underbrace{\left[ {\begin{array}{*{20}{c}}
p_{1,1}\mathbf{I}_N&p_{1,2}\mathbf{I}_N&\cdots&p_{1,K}\mathbf{I}_N\\
p_{2,1}\mathbf{I}_N&p_{2,2}\mathbf{I}_N&\cdots&p_{2,K}\mathbf{I}_N\\
\vdots&\vdots&\ddots&\vdots\\
p_{L_t,1}\mathbf{I}_N&p_{L_t,2}\mathbf{I}_N&\cdots&p_{L_t,K}\mathbf{I}_N
\end{array}} \right]}_{\mathbf{P}_s}\!
\underbrace{\left[ {\begin{array}{*{20}{c}}
\alpha_1\mathbf{h}_1\\
\alpha_2\mathbf{h}_2\\
\vdots\\
\alpha_K\mathbf{h}_K
\end{array}} \right]}_{\mathbf{h}_s}\!\!+\mathbf{w}\\
&\overset{(a)}{=}\overline{\mathbf{P}}
\underbrace{\left[ {\begin{array}{*{20}{c}}
\alpha_1\mathbf{\overline{h}}_1\\
\alpha_2\mathbf{\overline{h}}_2\\
\vdots\\
\alpha_K\mathbf{\overline{h}}_K
\end{array}} \right]}_{\mathbf{\overline{h}}}+\mathbf{w},
\end{split}
\end{equation}
where $\mathbf{P}_s=\mathbf{P}\otimes\mathbf{I}_N$, $\otimes$ represents the Kronecker product, $\mathbf{h}_s=vec(\mathbf{H}^T)$, and $\mathbf{h}_k$ is the effective channel gain vector of user $k$ on $N$ sub-carriers. According to the LDS spreading matrix $\mathbf{S}$, the transmitted sequence of each user is only spread onto $d_c$ sub-carriers. Therefore, as explained in (\ref{effvhannel}), $N-d_c$ elements in $\mathbf{h}_k$ are zero, and the positions of the zero elements are known to the BS. We further simplify $\mathbf{h}_s$ by eliminating the zeros according to $\mathbf{S}$. Accordingly, the columns in $\mathbf{P}_s$ corresponding to the zeros in $\mathbf{h}_s$ are also removed. Finally, we obtain the simplified version of (\ref{Pilotreceived}) in equation ($a$) of (\ref{Vectorization}), where $\overline{\mathbf{P}}$ and $\mathbf{\overline{h}}_k$ are obtained from $\mathbf{P}_s$ and $\mathbf{h}_k$ by zero elimination. According to (\ref{Vectorization}), the joint UAD and CE problem is formulated as
\begin{equation}\label{form}
\begin{split}
&\text{Solve: detect }\alpha_k\text{ and estimate }\overline{\mathbf{h}}_k\text{ if }\alpha_k\text{ is detected as }1\text{, }\forall k\\
&\text{Given:}\ \mathbf{y}\ \text{and}\ \overline{\mathbf{P}}
\end{split}
\end{equation}
\begin{remark}
To enable this joint UAD and CE task, each active user transmits its unique pilot sequence before its data transmission. A common practice is to employ the Zadoff-Chu (ZC) sequences \cite{ZC} as pilot sequences, which is also considered in this work. One ZC sequence with length $L_t$ is generated as follows,
\begin{equation}\label{ZC}
\mathbf{z}_{u}(n)=\exp\{-i\pi un(n+1)/L_t\},\ \ 0\leq n<L_t,
\end{equation}
where $i^2=-1$, the sequence length $L_t$ is a prime number, and $u \in [1,...,L_t-1]$ is the root of $\mathbf{z}_{u}(n)$. When the root $u$ is fixed, we can generate $N_u=\lfloor L_t/N_{SF}\rfloor$ different pilot sequences by cyclically shifting one existing sequence by $N_{SF}$ elements. Therefore, when $N_{SF}$ is set to 1, $L_t(L_t-1)$ different pilot sequences can be generated for the users.
\end{remark}
\subsection{MP-BSBL Algorithm \cite{MPBSBL}}
The recovery of the block sparse signal $\mathbf{\overline{h}}$ in (\ref{Vectorization}) can be addressed by the MP-BSBL algorithm \cite{MPBSBL}, which is based on two types of low-complexity message passing algorithms. The Gaussian message passing is employed for updating the estimated channel gain and the mean field (MF) message passing is adopted to approximate the hyper-parameters related to the user activity. The assumptions employed by the MP-BSBL algorithm are detailed as follows.
\begin{figure}
\centering
\includegraphics[width=1\linewidth]{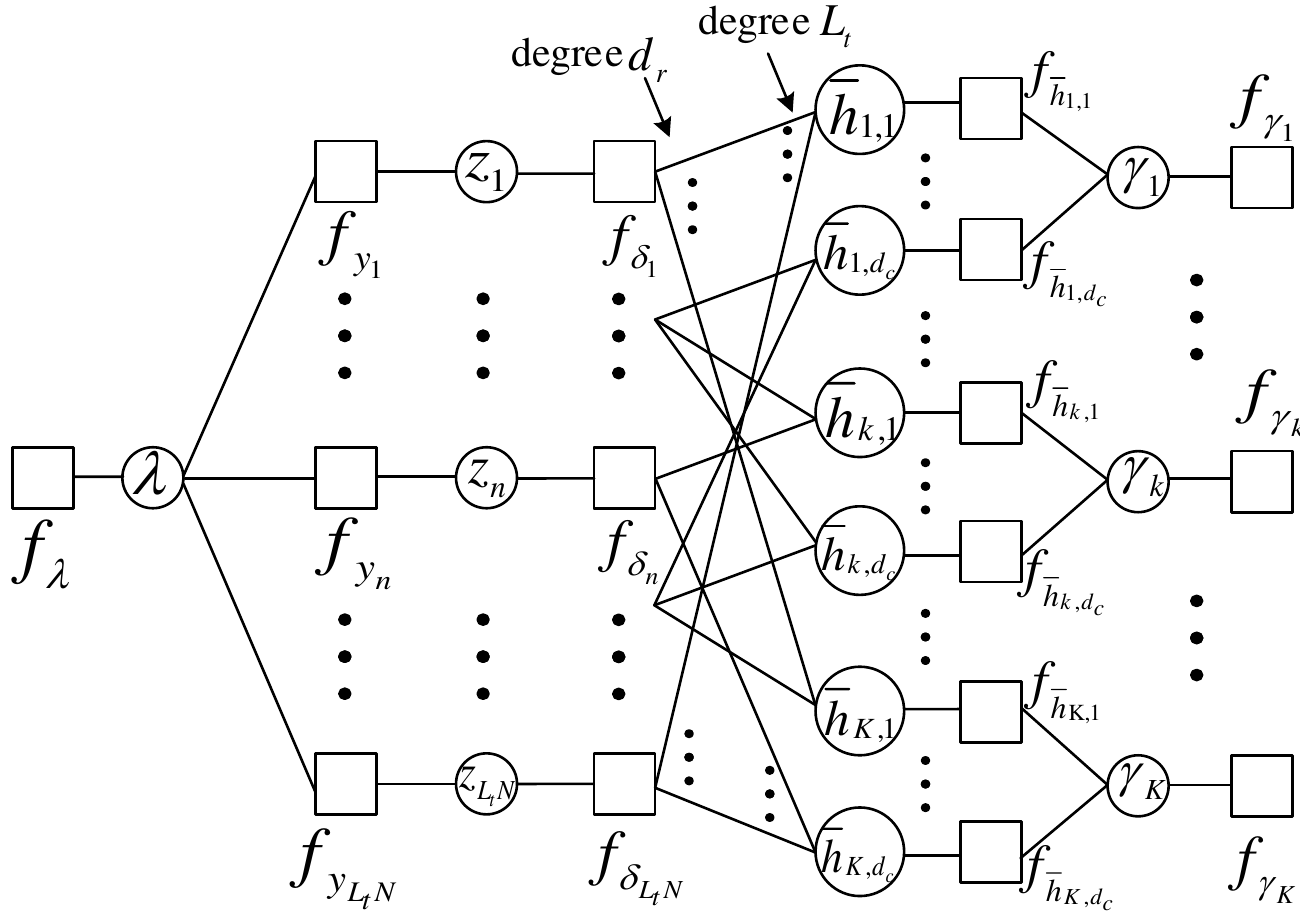}
\caption{Factor graph for the message passing in the MP-BSBL algorithm \cite{MPBSBL}.}
\label{FactorGraph}
\end{figure}

For user $k$, the distribution of the channel gains $\mathbf{\overline{h}}_k$ on $d_c$ sub-carriers is assumed conditioned on a hyper-parameter ${\gamma}_k$. Specifically, the channel gains are assumed independently and identically distributed (i.i.d) with complex Gaussian distribution, i.e., ${\mathbf{\overline{h}}_k}({\gamma}_k)\sim \mathcal{CN}(\mathbf{0},{\gamma}_k^{-1}\mathbf{I}_{d_c})$. In addition, the reciprocal of the estimated hyper-parameter is compared with a pre-defined threshold $\gamma_{th}$ to determine the user activity. If $\hat{{\gamma}}_k^{-1}$ is smaller than $\gamma_{th}$, this user is detected as inactive. Otherwise, this user is detected as active. The hyper-parameter ${\gamma}_k$ is assumed to follow a Gamma distribution \cite{MPBSBL2019,MPBSBL}. The noise precision $\lambda = 1/{\sigma_w^2}$ is unknown at the receiver but assumed with a prior probability $p(\lambda)$. With these assumptions above, the joint a posterior probability is factorized as follows
\begin{equation}\label{Factorize}
\begin{split}
p(\mathbf{\overline{h}},\gamma,\lambda|\mathbf{y}) &\propto p(\mathbf{y}|\mathbf{\overline{h}},\lambda)p(\mathbf{\overline{h}}|\mathbf{\gamma})p(\lambda)p(\gamma)\\
&=p(\lambda)\prod_{n=1}^{L_tN}p\left(y_n|\mathbf{\overline{h}},\lambda\right)\prod_{k=1}^{K}\prod_{d=1}^{d_c}p(\overline{h}_{k,d}|\gamma_k)p(\gamma_k),
\end{split}
\end{equation}
where $p(\lambda) \propto 1/\lambda$, $p(y_n|\mathbf{\overline{h}},\lambda)=\mathcal{CN}(y_n;\mathbf{\overline{p}^T_n}\mathbf{\overline{h}},\lambda^{-1})$, $\mathbf{\overline{p}^T_n}$ is the $n$-th row of matrix $\overline{\mathbf{P}}$ in (\ref{Vectorization}),  $p(\overline{h}_{k,d}|\gamma_k)=\mathcal{CN}(\overline{h}_{k,d};0,\gamma_k^{-1})$, $p(\gamma_k)=Ga(\gamma_k;a_k,b_k)$, and $p(\gamma)=\prod_{k=1}^{d_c}p(\gamma_k)$, $\mathcal{CN}(x;\mu,\sigma^2)$ represents the complex Gaussian distribution probability density function (pdf) of $x$ with mean $\mu$ and variance $\sigma^2$, while $Ga(x;a,b)$ represents the Gamma distribution pdf of $x$ with parameters $a$ and $b$. According to the variational Bayesian inference theory, the parameters $a$ and $b$ are usually assumed in the order of $10^{-4}$. Further details are referred to \cite{MPBSBL2019,MPBSBL}.

Accoring to the factorization in (\ref{Factorize}), a factor graph can be established for the MP-BSBL algorithm, which is illustrated in Fig. \ref{FactorGraph}. In Fig. 2, $f(\lambda)$, $f_{\overline{h}_{k,d}}(\overline{h}_{k,d},\gamma_k)$, $f_{\gamma_k}(\gamma_k)$ and $f_{y_n}(\mathbf{\overline{h}},\lambda)$ denote $p(\lambda)$, $p(\overline{h}_{k,d}|\gamma_k)$, $p(\gamma_k)$, and $p({y_n}|\mathbf{\overline{h}},\lambda)$, respectively. In addition, the extra variable $z_n=\mathbf{\overline{p}^T_n}\mathbf{\overline{h}}$ is introduced and the constraint $\delta(z_n-\mathbf{\overline{p}^T_n}\mathbf{\overline{h}})$ is represented by $f_{\delta_n}$. Then, $f_{y_n}$ is a function of $z_n$ and $\lambda$, i.e., $f_{y_n}(z_n,\lambda)=\mathcal{CN}(y_n;z_n,\lambda^{-1})$. The MP-BSBL algorithm performed on the factor graph in Fig. \ref{FactorGraph} is briefed as follows.

Denote $l$ as the iteration index and $Q_{k,d}$ as the product of all the incoming messages from the neighboring nodes $\delta_{n'}$ to $\overline{h}_{k,d}$, then according to $Q_{k,d}$, the variance $v_{Q_{k,d}}$ and mean $m_{Q_{k,d}}$ of $\overline{h}_{k,d}$ are derived as follows,
\begin{equation}\label{vQkd}
v^{l}_{Q_{k,d}}\approx\left( \sum_{n'\in\mathcal{N}\left(\overline{h}_{k,d}\right)}\frac{\left|\overline{P}_{n',kd}\right|^2}{\left(\hat{\lambda}^{l-1}\right)^{-1}+v^{l-1}_{\delta_{n'} \to z_{n'}}} \right)^{-1}
\end{equation}
\begin{equation}\label{mQkd}
m^{l}_{Q_{k,d}}\approx v^{l}_{Q_{k,d}} \sum_{n'\in\mathcal{N}\left(\overline{h}_{k,d}\right)}\frac{\overline{P}_{n',kd}^{H}\left(y_{n'}-m^{l-1}_{\delta_{n'} \to z_{n'}}\right)}{\left(\hat{\lambda}^{l-1}\right)^{-1}+v^{l-1}_{\delta_{n'} \to z_{n'}}}+m_{\overline{h}_{k,d}}^{l-1}
\end{equation}

By considering the message $\hat{\gamma}_k^{l-1}$ passing from $f_{\overline{h}_{k,d}}$ to $\overline{h}_{k,d}$, the variance $v^{l}_{\overline{h}_{k,d}}$ and mean $m^{l}_{\overline{h}_{k,d}}$ of $\overline{h}_{k,d}$ are updated as
\begin{equation}\label{vmhkd}
\begin{split}
v^{l}_{\overline{h}_{k,d}}&=\frac{1}{\left(v_{Q_{k,d}}^{l}\right)^{-1}+\hat{\gamma}_k^{l-1}}\\
m^{l}_{\overline{h}_{k,d}}&=\frac{m^{l}_{Q_{k,d}}}{1+v^{l}_{Q_{k,d}}\hat{\gamma}_k^{l-1}}
\end{split}
\end{equation}

The variance $v_{\delta_n \to z_n}$ and mean $m_{\delta_n \to z_n}$ passing from $\delta_n$ to $z_n$ are derived in (\ref{delta2z}).
\begin{equation}\label{delta2z}
\begin{split}
v^{l}_{\delta_n \to z_n}\! &\approx \!\sum_{ \left\{ i,j\right\} \in \mathcal{N}\left(f_{\delta_n}\right)  }\left| \overline{P}_{n,ij}\right|^2v^{l}_{\overline{h}_{i,j}}\\
m^{l}_{\delta_n \to z_n}\! &\approx \!\sum_{ \left\{ i,j\right\} \in \mathcal{N}\left(f_{\delta_n}\right)  }\overline{P}_{n,ij}m^{l}_{\overline{h}_{i,j}}\!-\!\frac{v^{l}_{\delta_n \to z_n}\left( y_n\!-\!m_{\delta_n \to z_n}^{l-1}\right)}{\left(\hat{\lambda}^{l-1}\right)^{-1}\!+\!v_{\delta_n \to z_n}^{l-1}}
\end{split}
\end{equation}

Then the hyper-parameter $\hat{\gamma}^{l}_k$ for the variance of  the channel gain is updated according to the MF message passing,
\begin{equation}\label{gamma}
\hat{\gamma}^{l}_k=\frac{a_k+d_c+1}{b_k+\sum\limits_{d=1}^{d_c}\left( \left|m^{l}_{\overline{h}_{k,d}}\right|^2+v^{l}_{\overline{h}_{k,d}}\right)}
\end{equation}

The variance $v^{l}_{z_n}$ and mean $m^{l}_{z_n}$ of $z_n$ are derived by $v_{\delta_n \to z_n}$, $m_{\delta_n \to z_n}$ and the noise precision $\hat{\lambda}^{l-1}$,
\begin{equation}\label{zn}
\begin{split}
v^{l}_{z_n}&=\left(\hat{\lambda}^{l-1}+\left(v^{l}_{\delta_n \to z_n}\right)^{-1}\right)^{-1}\\
m^{l}_{z_n}&=v^{l}_{z_n}\left(y_n\hat{\lambda}^{l-1}+\frac{m^{l}_{\delta_n \to z_n}}{v^{l}_{\delta_n \to z_n}}\right)
\end{split}
\end{equation}

Finally, the noise precision $\hat{\lambda}^{l}$ is updated with the MF message passing,
\begin{equation}\label{lambda}
\hat{\lambda}^{l}=\frac{L_tN}{\sum\limits_{n=1}^{L_tN}\left[ \left(m^{l}_{z_n}-y_n\right)^2+v^{l}_{z_n}\right]}
\end{equation}

At the last iteration, if $\left({\hat{\gamma}^l_k}\right)^{-1}$ is smaller than a predefined threshold $\gamma_{th}$, user $k$ is regarded as inactive. Otherwise, this user is determined as active and  $\{m^l_{\overline{h}_{k,d}}, d=1,\ldots,d_c\}$ is the estimated channel gain for user $k$. Further details of the derivations above are referred to \cite{MPBSBL}.
\section{DNN-Aided MP-BSBL Algorithm}\label{DNN_SBSL}
\begin{figure*}
\centering
\includegraphics[width=1.0\linewidth]{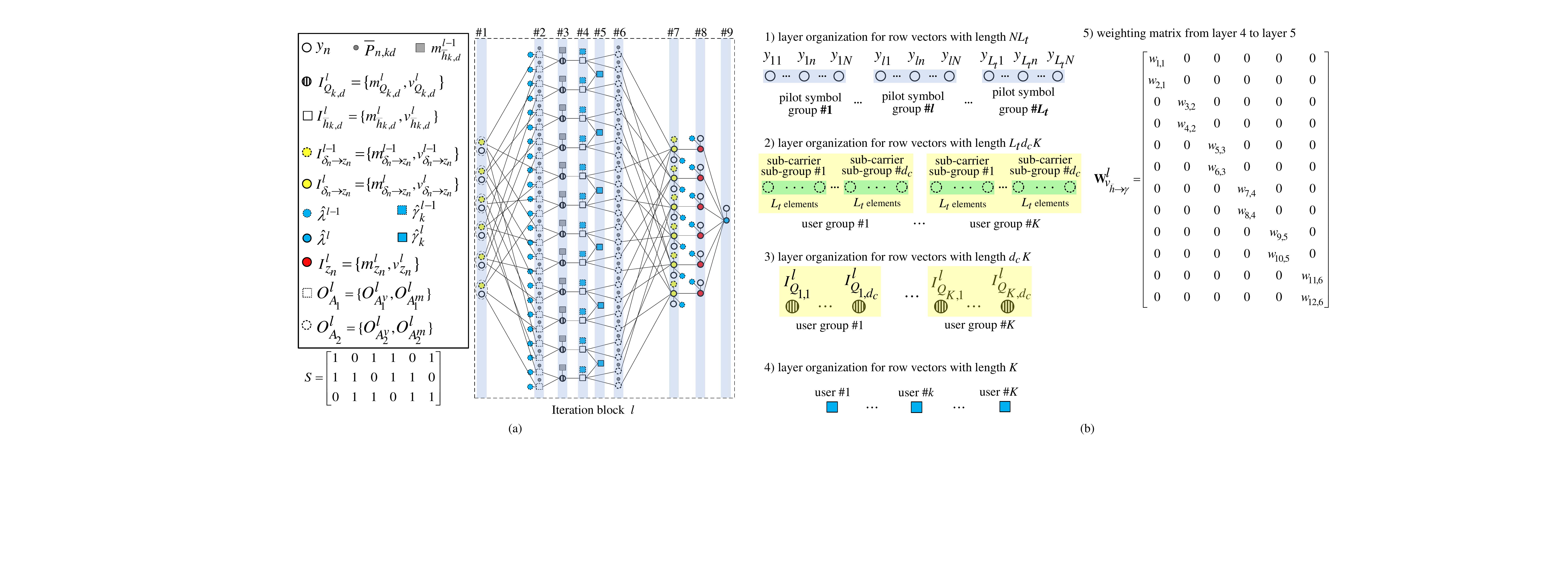}\vspace{-0.3cm}
\caption{DNN for the weighted message passing in DNN-MP-BSBL algorithm with ($N=3, K=6, L_t=2, d_c=2$) and layer organization.}\vspace{-0.3cm}
\label{DNN-MP-BSBL}
\end{figure*}
Two types of message passing are adopted by the MP-BSBL algorithm. The Gaussian message passing is employed for CE in (\ref{vmhkd}), while the MF message passing is employed to update the hyper-parameters $\hat{\gamma}_k^l$ in (\ref{gamma}) for UAD. According to the factor graph in Fig. \ref{FactorGraph}, the variable nodes $\overline{h}_{k,d}$ and sum nodes $f_{\delta_n}$ are densely connected. That is, the connection between $\overline{h}_{k,d}$ and $f_{\delta_n}$ is characterized by many short cycles with girth 4, which results in the correlation problem of the Gaussian messages \cite{GMPID}. Furthermore, in crowded NORA systems, where the number of sum nodes $f_{\delta_n}$ is much smaller than the number of variable nodes $\overline{h}_{k,d}$, the correlated Gaussian messages further lead to the convergence problem for the MP-BSBL algorithm. As a result, the convergence speed is deteriorated and the accuracy of the MP-BSBL algorithm fails to get improved as the signal-to-noise ratio (SNR) increases. On the other hand, the correlated Gaussian messages also lead to inaccuracy for the MF message update, which undermines the UAD performance.

Confronted with the problems above, we propose a DNN-MP-BSBL algorithm, which imposes weights on the Gaussian messages on the factor graph in Fig. \ref{FactorGraph}. Simultaneously, weights are also imposed on the MF message update in (\ref{gamma}) and further trained to improve the UAD accuracy. To facilitate the training of the weights, we transfer the message passing process from the factor graph in Fig. \ref{FactorGraph} to a DNN structure. The back-and-forth message passing process on the factor graph is now transformed into a forward-propagation trellis in Fig. \ref{DNN-MP-BSBL}(a), which resembles the structure of a neural network. It is assumed that there are $N=3$ sub-carriers, $K=6$ users, the length of pilot sequence is $L_t=2$, each user chooses $d_c=2$ sub-carriers to perform uplink data transmission, and the DNN-MP-BSBL algorithm is performed with $N_{it}$ iterations. The regular LDS spreading matrix $\mathbf{S}$ is also shown in Fig. \ref{DNN-MP-BSBL}(a).

As shown in Fig. \ref{DNN-MP-BSBL}(a), the trellis can be considered as a neural network. Each iteration of the MP-BSBL algorithm is now represented by one iteration block. Since the structure is identical for each iteration block, we only illustrate one iteration block in Fig. \ref{DNN-MP-BSBL}(a). However, it is noted that neighboring iteration blocks are connected. For illustration simplicity, we include the nodes from the $(l-1)$-th iteration block, e.g. $I^{l-1}_{\delta_n\to z_n}$ and $\hat{\lambda}^{l-1}$ as the input nodes in the $l$-th iteration block. In this way, the entire DNN is constructed by connecting $N_{it}$ iteration blocks. Within each iteration block, one layer represents one particular message, i.e., each layer is defined by its output message. Two auxiliary layers $A_1$ and $A_2$, i.e., Layer 2 and Layer 6 are also added for the clarity of illustration. Therefore, according to equations (\ref{vQkd}) to (\ref{lambda}), there are 9 layers within each iteration block. Detailed organization of each layer is illustrated in Fig. \ref{DNN-MP-BSBL}(b) and the weighted message passing process is explained as follows.
\begin{table}
\renewcommand\arraystretch{1.5}
\caption{Input and Output for Each Layer of the DNN.}
\centering
\begin{tabular}{cccc}
  \Xhline{1.2pt}
\!\!\!\!Index&\!\!\!\!\!\!\!\!\!\!\!\!\!\!Layer Input&\!\!Layer Output&\!\!\!\!Length\!\!\\
  \hline
\!\!\!\!1&\!\!\!\!\!\!\!\!\!\!\!\!\!\!None&\!\!\!\!$\mathbf{y},\mathbf{I}^{l-1}_{\delta \!\to\! z}\!=\!\{\mathbf{v}^{l-1}_{\delta \!\to\! z},\mathbf{m}^{l-1}_{\delta \!\to\! z}\}$&\!\!\!\!$NL_t$\!\!\\
\!\!\!\!2&\!\!\!\!\!\!\!\!\!\!$\mathbf{y},\mathbf{I}^{l-1}_{\delta \!\to\! z},\hat{\lambda}^{l-1},\mathbf{P}$&\!\!\!\!$\mathbf{O}^l_{A_1}\!=\!\{\mathbf{O}^l_{A^v_1},\mathbf{O}^l_{A^m_1}\}$&\!\!\!\!$L_td_cK$\!\!\\
\!\!\!\!3&\!\!\!\!\!\!\!\!\!\!$\mathbf{m}^{l-1}_{\overline{h}},\mathbf{O}^l_{A_1}$&\!\!\!\!$\mathbf{I}^l_Q\!=\!\{\mathbf{v}^l_Q,\mathbf{m}^l_Q\}$&\!\!\!\!$d_cK$\!\!\\
\!\!\!\!4&\!\!\!\!\!\!\!\!\!\!$\mathbf{I}^l_Q,\hat{\gamma}^{l-1}$&\!\!\!\!$\mathbf{I}^l_{\overline{h}}\!=\!\{\mathbf{v}^l_{\overline{h}},\mathbf{m}^l_{\overline{h}}\}$&\!\!\!\!$d_cK$\!\!\\
\!\!\!\!5&\!\!\!\!\!\!\!\!\!\!$\mathbf{I}^l_{\overline{h}}$&\!\!\!\!$\hat{\mathbf{\gamma}}^l$&\!\!\!\!$K$\!\!\\
\!\!\!\!6&\!\!\!\!\!\!\!\!\!\!$\mathbf{P},\mathbf{I}^l_{\overline{h}}$&\!\!\!\!$\mathbf{O}^l_{A_2}\!=\!\{\mathbf{O}^l_{A^v_2},\mathbf{O}^l_{A^m_2}\}$&\!\!\!\!$L_td_cK$\!\!\\
\!\!\!\!7&\!\!\!\!\!\!\!\!\!\!$\mathbf{y},\mathbf{I}^{l-1}_{\delta \!\to\! z},\hat{\lambda}^{l-1},\mathbf{O}^l_{A_2}$&\!\!\!\!$\mathbf{I}^l_{\delta \!\to\! z}\!=\!\{\mathbf{v}^l_{\delta \!\to\! z},\mathbf{m}^l_{\delta \!\to\! z}\}$&\!\!\!\!$NL_t$\!\!\\
\!\!\!\!8&\!\!\!\!\!\!\!\!\!\!$\mathbf{y},\hat{\lambda}^{l-1},\mathbf{I}^l_{\delta \!\to\! z}$&\!\!\!\!$\mathbf{I}^l_{z}\!=\!\{\mathbf{v}^l_{z},\mathbf{m}^l_{z}\}$&\!\!\!\!$NL_t$\!\!\\
\!\!\!\!9&\!\!\!\!\!\!\!\!\!\!$\mathbf{I}^l_{z},\mathbf{y}$&\!\!\!\!$\hat{\lambda}^l$&\!\!\!\!$1$\!\!\\
  \Xhline{1.2pt}
\end{tabular}
\label{LayerMeaning}
\end{table}
\subsection{Layer Organization and Weighting Matrix}
For notational simplicity, the nodes in each layer are placed in a row vector in Fig. \ref{DNN-MP-BSBL}(b), and the weighted message passing process is then equivalent to the input vector multiplying a weighting matrix. The input vectors, output vector, and the length of output vector for each layer are listed in Table \ref{LayerMeaning} while the elements of different vectors are organized as follows.
\subsubsection{For vectors with length $NL_t$} The organization for the row vector $\mathbf{y}$ in Layer 1 is taken as an example in Fig. \ref{DNN-MP-BSBL}(b). Every $N$ elements are grouped into a pilot symbol group according to the common pilot symbol they correspond to. These $L_t$ pilot symbol groups are organized according to the index of the pilot symbols. Within each pilot symbol group, the $N$ elements are organized according to the sub-carrier index. For example, the $n$-th node $y_{ln}$ in the $l$-th pilot symbol group represents the $l$-th received pilot signal on the $n$-th sub-carrier. In this way, the row vector $\mathbf{y}$ is organized as $\mathbf{y}=[y_{11},\ldots,y_{1N},\ldots,y_{L_t1},\ldots,y_{L_tN}]$.
\subsubsection{For vectors with length $L_td_cK$} The organization for the auxiliary row vector $\mathbf{O}^l_{A_2}$ in Layer 6 is taken as an example. Every $L_td_c$ elements are grouped into a user group, according to the common user they correspond to. Within each user group, every $L_t$ elements are grouped into a sub-group according to the common sub-carrier they correspond to. Within each sub-carrier sub-group, the $L_t$ elements are arranged according to the pilot symbol index.
\subsubsection{For vectors with length $d_cK$} The organization for the row vector $\mathbf{I}^l_{Q}$ in Layer 3 is taken as an example. Every $d_c$ elements are grouped together according to the common user they correspond to. Within each user group, the $d_c$ elements are organized according to the sub-carrier index. In this way, the row vector $\mathbf{I}^l_{Q}$ is organized as $\mathbf{I}^l_{Q}=[I^l_{Q_{1,1}},\ldots,I^l_{Q_{1,d_c}},\ldots,I^l_{Q_{K,1}},\ldots,I^l_{Q_{K,d_c}}]$, where $I^l_{Q_{k,d}}=\{m^l_{Q_{k,d}},v^l_{Q_{k,d}}\}$.
\subsubsection{For vectors with length $K$} The $K$ elements in the row vector $\hat{\mathbf{\gamma}}^l$ are arranged according to the user index.
\subsubsection{Weighting matrix} Since input vectors and output vector of a layer are organized as row vectors, the weighted message update (i.e., the output vector) for every layer can be represented by the input vector multiplied by a weighting matrix. The size of the weighting matrix is $L_I\times L_O$ where $L_I$ and $L_O$ are the length of the input vector and the output vector, respectively. According to Fig. \ref{DNN-MP-BSBL}(a), nodes in neighboring layers are not fully connected, i.e., the weighting matrix is defined by the inter-layer connection. An example is shown in Fig. \ref{DNN-MP-BSBL}(b) for the weighting matrix defined by the connection from Layer 4 to Layer 5 of Fig. \ref{DNN-MP-BSBL}(a). According to this example, the $(i,j)$-th entry of the weighting matrix is non-zero if and only if the $i$-th node in the input layer is connected to the $j$-th node in the output layer. Furthermore, only the non-zero weights in the weighting matrix are trained for the DNN-MP-BSBL algorithm.
\subsection{Weighted Message Passing}
The weighted message passing is explained as follows.

\noindent\underline{\textbf{Layer 1}}: Layer 1 serves as the input of the neural network within one iteration block and it is composed of two message vectors,  $\mathbf{y}=\{y_n, \forall n\}$ and $\mathbf{I}^{l-1}_{\delta \to z}=\{\mathbf{v}^{l-1}_{\delta \to z},\mathbf{m}^{l-1}_{\delta \to z}\}=\{v^{l-1}_{\delta_n \to z_n},m^{l-1}_{\delta_n \to z_n}, \forall n\}$. The elements $y_n$ and $I^{l-1}_{\delta_n \to z_n}=\{v^{l-1}_{\delta_n \to z_n},m^{l-1}_{\delta_n \to z_n}\}$ are grouped together in Fig. \ref{DNN-MP-BSBL}(a) since they share identical connection to the nodes in Layer 2.

\noindent\underline{\textbf{Layer 2}}: Layer 2 is the auxiliary layer $A^l_1$, and the output $\mathbf{O}^l_{A_1}$ is composed of two message vectors $\mathbf{O}^l_{A^v_1}$ and $\mathbf{O}^l_{A^m_1}$, whose elements correspond to the terms in the cumulative summation of (\ref{vQkd}) and (\ref{mQkd}), respectively. The weighted message update of Layer 2 is derived as follows,
\begin{equation}\label{W_OAv}
\mathbf{O}^l_{A^v_1}=\frac{\mathbf{P}^2}  {\frac{1}{\hat{\lambda}^{l-1}}\times \mathbf{W}^{l}_{\lambda\to A^v_1}+\mathbf{v}^{l-1}_{\delta \to z}\times\mathbf{W}^l_{v_\delta\to A^v_1}}
\end{equation}
\begin{equation}\label{W_OAm}
\mathbf{O}^l_{A^m_1}=\frac{\mathbf{P}^H.*\left( \mathbf{y}\times\mathbf{W}^{l}_{y\to A^m_1} -\mathbf{m}^{l-1}_{\delta\to z}\times\mathbf{W}^{l}_{{m_\delta}\to A^m_1}  \right)} {\frac{1}{\hat{\lambda}^{l-1}}\times \mathbf{W}^{l}_{\lambda\to A^m_1}+\mathbf{v}^{l-1}_{\delta \to z}\times\mathbf{W}^l_{v_\delta\to A^m_1}}
\end{equation}
where $\mathbf{P}^2=\{|\overline{P}_{n,kd}|^2, \forall n,k,d\}$ and $\mathbf{P}^H=\{\overline{P}^H_{n,kd}, \forall n,k,d\}$ are row vectors with length $L_td_cK$. $\mathbf{W}^{l}_{\lambda\to A^v_1}$ and $\mathbf{W}^{l}_{\lambda\to A^m_1}$ are two weighting matrices with size $1\times L_td_cK$. $\mathbf{W}^l_{v_\delta\to A^v_1}$, $\mathbf{W}^{l}_{y\to A^m_1}$, $\mathbf{W}^{l}_{{m_\delta}\to A^m_1}$, and $\mathbf{W}^l_{v_\delta\to A^m_1}$ are weighting matrices defined by the connection from Layer 1 to Layer 2. In addition, the fraction and $.*$ operations are the element-wise division and multiplication operations while the $\times$ operation refers to the matrix multiplication operation.

\noindent\underline{\textbf{Layer 3}}: The weighted output message vectors $\mathbf{I}^l_Q=\{\mathbf{v}^l_Q,\mathbf{m}^l_Q\}=\{v^l_{Q_{k,d}},m^l_{Q_{k,d}},\forall k,d\}$ of Layer 3 are derived as follows,
\begin{equation}\label{W_Q}
\begin{split}
\mathbf{v}^l_Q&=\frac{1}{\mathbf{O}^l_{A^v_1}\times\mathbf{W}^{l}_{A^v_1\to v_Q}}\\
\mathbf{m}^l_Q&=\mathbf{v}^l_Q.*\left( \mathbf{O}^l_{A^m_1} \times \mathbf{W}^{l}_{A^m_1 \to m_Q} \right)+\mathbf{m}^{l-1}_{\overline{h}} \times \mathbf{W}^l_{\overline{h} \to Q}
\end{split}
\end{equation}
where $\mathbf{W}^{l}_{A^v_1\to v_Q}$ and $\mathbf{W}^{l}_{A^m_1\to m_Q}$ are weighting matrices defined by the connection from Layer 2 to Layer 3. $\mathbf{W}^l_{\overline{h}\to Q}$ is a diagonal weighting matrix with size $d_cK\times d_cK$.

\noindent\underline{\textbf{Layer 4}}: The weighted output message vectors $\mathbf{I}^l_{\overline{h}}=\{\mathbf{v}^l_{\overline{h}}, \mathbf{m}^l_{\overline{h}}\}=\{v^l_{\overline{h}_{k,d}}, m^l_{\overline{h}_{k,d}}, \forall k,d\}$ of Layer 4 are derived as follows,
\begin{equation}\label{W_h}
\begin{split}
\mathbf{v}^l_{\overline{h}}&=\frac{1}{\frac{1}{\mathbf{v}_Q^l} + \left(\mathbf{\hat{\gamma}}^{l-1}\bigotimes \mathbf{1}_{d_c}\right) \times \mathbf{W}^l_{\gamma}}\\
\mathbf{m}^l_{\overline{h}}&=\frac{\mathbf{m}^l_Q}{\mathbf{1}_{d_cK}\times \mathbf{W}_{\text{one}\to\overline{h}} + \left(\mathbf{v}^l_Q .*(\mathbf{\hat{\gamma}}^{l-1}\bigotimes \mathbf{1}_{d_c})\right) \times \mathbf{W}^l_{v\gamma \to \overline{h}}}
\end{split}
\end{equation}
where $\bigotimes$ refers to the Kronecker product. $\mathbf{1}_{d_c}$ and $\mathbf{1}_{d_cK}$ are two all-one row vectors and their vector length are $d_c$ and $d_cK$, respectively. $\mathbf{W}^l_{\gamma}$, $\mathbf{W}_{\text{one}\to\overline{h}}$, and $\mathbf{W}^l_{v\gamma \to \overline{h}}$ are diagonal weighting matrices with size $d_cK\times d_cK$.

\noindent\underline{\textbf{Layer 5}}: The weighted output message vector $\hat{\gamma}^l=\{\hat{\gamma}^l_k, \forall k\}$ is derived as follows,
\begin{equation}\label{W_gamma}
\mathbf{\hat{\gamma}}^l=\frac{\mathbf{a}+d_c+1}{\mathbf{b}+|\mathbf{m}^l_{\overline{h}}|^2\times \mathbf{W}^l_{m_{\overline{h}}\to\gamma} + \mathbf{v}^l_{\overline{h}} \times \mathbf{W}^l_{v_{\overline{h}}\to\gamma}}
\end{equation}
where $\mathbf{a}=\{a_k, \forall k\}$ and $\mathbf{b}=\{b_k, \forall k\}$ are row vectors with length $K$, $|\mathbf{m}^l_{\overline{h}}|^2=\{|m^l_{\overline{h}_{k,d}}|^2, \forall k,d\}$, $\mathbf{W}^l_{m_{\overline{h}}\to\gamma}$ and $\mathbf{W}^l_{v_{\overline{h}}\to\gamma}$ are weighting matrices defined by the connection from Layer 4 to Layer 5.

\noindent\underline{\textbf{Layer 6}}: Layer 6 is the auxiliary layer $A_2^l$ with weighted output message vectors $\mathbf{O}^l_{A^v_2}$ and $\mathbf{O}^l_{A^m_2}$, whose elements correspond to the terms in the cumulative summation in (\ref{delta2z}).
\begin{equation}\label{OA2}
\begin{split}
\mathbf{O}^l_{A^v_2}&=\mathbf{P}^2.*\mathbf{v}^l_{\overline{h}}\\
\mathbf{O}^l_{A^m_2}&=\mathbf{P}.*\mathbf{m}^l_{\overline{h}}
\end{split}
\end{equation}
where $\mathbf{P}=\{\overline{P}_{n,kd}, \forall n,k,d\}$ is a row vector with length $L_td_cK$.

\noindent\underline{\textbf{Layer 7}}: The weighted output vectors $\mathbf{I}^l_{\delta\to z}=\{\mathbf{v}^l_{\delta\to z},\mathbf{m}^l_{\delta\to z}\}=\{{v}^l_{\delta_n\to z_n},{m}^l_{\delta_n\to z_n}, \forall n\}$ of Layer 7 are derived as follows,
\begin{equation}\label{W_delta2z}
\begin{split}
\mathbf{v}^l_{\delta\to z}&=\mathbf{O}^l_{A^v_2} \times \mathbf{W}^l_{A_2^v \to v_\delta}\\
\mathbf{m}^l_{\delta\to z}&=\mathbf{O}^l_{A^m_2} \times \mathbf{W}^l_{A_2^m \to m_\delta}\\
&-\frac{\mathbf{v}_{\delta\to z}.*\left(\mathbf{y}\times \mathbf{W}^l_{y\to \delta}-\mathbf{m}^{l-1}_{\delta\to z}\times \mathbf{W}^l_{m_\delta \to m_\delta} \right)}{\left(\hat{\lambda}^{l-1}\right)^{-1}\times \mathbf{W}^l_{\lambda \to \delta}+\mathbf{v}^{l-1}_{\delta\to z}\times \mathbf{W}^{l}_{v_\delta \to m_\delta}}
\end{split}
\end{equation}
where $\mathbf{W}^l_{A_2^v \to v_\delta}$ and $\mathbf{W}^l_{A_2^m \to m_\delta}$ are weighting matrices defined by the connection from Layer 6 to Layer 7. $\mathbf{W}^l_{y\to \delta}, \mathbf{W}^l_{m_\delta \to m_\delta}$ and $\mathbf{W}^{l}_{v_\delta \to m_\delta}$ are diagonal weighting matrices with size $NL_t\times NL_t$. $\mathbf{W}^l_{\lambda \to \delta}$ is a weighting matrix with size $1\times NL_t$.

\noindent\underline{\textbf{Layer 8}}: The weighted output vectors $\mathbf{I}^l_{z}=\{\mathbf{v}^l_{z},\mathbf{m}^l_{z}\}=\{{v}^l_{z_n},{m}^l_{z_n}, \forall n\}$ of Layer 8 are derived as follows,
\begin{equation}\label{Wz}
\begin{split}
\mathbf{v}^l_z&=\frac{1}{\hat{\lambda}^{l-1}\times\mathbf{W}^l_{\lambda\to z}+\frac{1}{\mathbf{v}^l_{\delta\to z}}\times\mathbf{W}_{{v_\delta}\to v_z}}\\
\mathbf{m}^l_z&=\mathbf{v}^l_z .*\left( (\mathbf{y}.*\hat{\lambda}^{l-1})\times \mathbf{W}^l_{y\lambda\to z} + \frac{\mathbf{m}^l_{\delta\to z}}{\mathbf{v}^l_{\delta\to z}}\times \mathbf{W}^l_{mv\to z}\right)
\end{split}
\end{equation}
where $\mathbf{W}_{{v_\delta}\to v_z}$ and $\mathbf{W}^l_{mv\to z}$ are weighting matrices defined by the connection from Layer 7 to Layer 8. $\mathbf{W}^l_{y\lambda\to z}$ is a diagonal weighting matrix with size $NL_t\times NL_t$, and $\mathbf{W}^l_{\lambda\to z}$ is a weighting matrix with size $1\times NL_t$.

\noindent\underline{\textbf{Layer 9}}: The weighted output message $\hat{\lambda}^l$ of Layer 9 is derived as follows,
\begin{equation}\label{W_lambda}
\hat{\lambda}^l\!=\!\frac{L_tN}{\left(\mathbf{m}^l_z \!\times\! \mathbf{W}^l_{m_z\to\lambda}\!-\!\mathbf{y}\!\times\!\mathbf{W}^l_{y\to\lambda}\right)^2\!\times\! \mathbf{1}^T_{N L_t}\!+\!\mathbf{v}^l_z\!\times\! \mathbf{W}^l_{v_z\to\lambda}}
\end{equation}
where $\mathbf{W}^l_{m_z\to\lambda}$ and $\mathbf{W}^l_{y\to\lambda}$ are diagonal weighting matrices with size $NL_t \times NL_t$, and $\mathbf{1}^T_{NL_t}$ is an all-one column vector with length $NL_t$. $\mathbf{W}^l_{v_z\to\lambda}$ is a weighting matrix with size $NL_t\times 1$.
\subsection{Loss Function}
The DNN-MP-BSBL algorithm is conducted in two periods: the training period and the testing period. In the training period, according to the given samples in the training set, the non-zero entries in the weighting matrices are trained to minimize the estimation error. In the testing period, the weighting matrices are fixed. The UAD and CE accuracy of the DNN-MP-BSBL algorithm is evaluated by the samples in the test set.

In order to train the non-zero entries in the weighting matrices, a loss function is employed during the training period to measure the estimation error. Then the weights are adjusted with the stochastic gradient descent (SGD) algorithm in a back-propagation manner to minimize the loss function. For the stability of the SGD algorithm, we employ the mean square error (MSE) $\|\hat{\mathbf{h}}_\text{DNN}-\mathbf{h}\|_2^2$ as the loss function, where $\hat{\mathbf{h}}_\text{DNN}$ is the estimated channel gain and $\mathbf{h}$ is the known channel gain in the training set. It is noted that, different from the training period, the Normalized MSE (NMSE) $\|\hat{\mathbf{h}}_\text{DNN}-\mathbf{h}\|_2^2/\|\mathbf{h}\|_2^2$ is considered for the simulations in the testing period to measure the UAD and CE accuracy.
\begin{algorithm}[t!]\setstretch{1.2}

               \caption{DNN-MP-BSBL algorithm}

               \label{alg:MP-BSBL}


               \KwIn{$\mathbf{y}$, $\mathbf{\overline{P}}$, $N_{it}$, $d_c$,}

               {\ \ \ \ \ \ \ \ trained weighting matrices in equations (\ref{W_OAv}) to (\ref{W_lambda})}

               \KwOut{$\hat{\mathbf{h}}_{\text{DNN}}$, the index set of active user IDX}

               {\textbf{Initialize:}}

               {\ \ \ \ $l=0;\hat{\lambda}^l=10^3;\mathbf{\hat{\gamma}}^l=\mathbf{1};$}

               {\ \ \ \ $\mathbf{v}^l_{\delta\to z}=\mathbf{1}, \mathbf{m}^l_{\delta\to z}=\mathbf{0};\mathbf{m}^l_{\overline{h}}=\mathbf{0}$.}

               \For{$l=1:N_{it}$}{

               1. With $\hat{\lambda}^{l-1},\mathbf{m}^{l-1}_{\overline{h}},\mathbf{v}^{l-1}_{\delta\to z}$, and $\mathbf{m}^{l-1}_{\delta\to z}$ in the ($l-1$)-th iteration, update $\mathbf{v}^l_Q$ and $\mathbf{m}^l_Q$ by (\ref{W_OAv}), (\ref{W_OAm}), and (\ref{W_Q}).

               2. With $\mathbf{v}^l_Q$, $\mathbf{m}^l_Q$ in the $l$-th iteration and $\mathbf{\hat{\gamma}}^{l-1}$ in the ($l-1$)-th iteration, update $\mathbf{v}^l_{\overline{h}}$ and $\mathbf{m}^l_{\overline{h}}$ by (\ref{W_h}).

               3. With $\mathbf{v}^l_{\overline{h}}$ and $\mathbf{m}^l_{\overline{h}}$ in the $l$-th iteration, update $\mathbf{\hat{\gamma}}^l$ by (\ref{W_gamma}).

               4. With $\hat{\lambda}^{l-1}, \mathbf{v}^{l-1}_{\delta\to z}$, $\mathbf{m}^{l-1}_{\delta\to z}$ in the ($l-1$)-th iteration and $\mathbf{v}^l_{\overline{h}}$, $\mathbf{m}^l_{\overline{h}}$ in the $l$-th iteration, update $\mathbf{v}^{l}_{\delta\to z}$ and $\mathbf{m}^{l}_{\delta\to z}$ by (\ref{OA2}) and (\ref{W_delta2z}).

               5. With $\hat{\lambda}^{l-1}$ in the ($l-1$)-th iteration and $\mathbf{v}^{l}_{\delta\to z}$, $\mathbf{m}^{l}_{\delta\to z}$ in the $l$-th iteration, update $\mathbf{v}^l_z$ and $\mathbf{m}^l_z$ by (\ref{Wz}).

               6. With $\mathbf{v}^l_z$ and $\mathbf{m}^l_z$ in the $l$-th iteration, update the noise precision $\hat{\lambda}^{l}$ by (\ref{W_lambda}).
               }
    \textbf{return:}

    {\ \ \ \ IDX=find$\left(\{\left({\hat{\gamma}^l_k}\right)^{-1}\}>\gamma_{th}\right)$,}

    {\ \ \ \ $\hat{\mathbf{h}}_{\text{DNN}}=\{\mathbf{0}, k \notin \text{IDX}\} \bigcup \{m^l_{\overline{h}_{k,d}}, k \in \text{IDX}, d=1,\ldots,d_c\}$.}
\end{algorithm}
\subsection{Summary of the Proposed DNN-MP-BSBL Algorithm}
After the training period, all the weighting matrices are fixed, and the proposed DNN-MP-BSBL algorithm is summarized in Algorithm \ref{alg:MP-BSBL}.  After $N_{it}$ iterations, the UAD decision is made by comparing $\left({\hat{\gamma}^{N_{it}}_k}\right)^{-1}$ with a threshold $\gamma_{th}$. If $\left({\hat{\gamma}^{N_{it}}_k}\right)^{-1}\leq\gamma_{th}$, user $k$ is detected as inactive. Otherwise, this user is detected as active and $\{m^{N_{it}}_{\overline{h}_{k,d}}, \forall d\}$ is the estimated channel gain for user $k$.
\begin{table}
\renewcommand\arraystretch{1.4}
\caption{Extra number of multiplications $\Delta_\times$ caused by weighting matrices in DNN-MP-BSBL algorithm.}
\centering
\begin{tabular}{c|c|c}
  \Xhline{1.2pt}
\!Weighting Matrix\!&\!Eqn\!&\!$\Delta_{\times}$\!\\\hline
\!\!\!$\mathbf{W}^{l}_{\lambda\!\to\! A^v_1},\!\mathbf{W}^{l}_{\lambda\!\to\! A^m_1},\!\mathbf{W}^l_{v_\delta\!\to\! A^v_1},\!\mathbf{W}^l_{v_\delta\!\to\! A^m_1}$\!\!\!&{(\ref{W_OAv}),(\ref{W_OAm})}&{$L_td_cK$}\\\hline
$\mathbf{W}^{l}_{y\to A^m_1},\mathbf{W}^{l}_{{m_\delta}\to A^m_1}$&{(\ref{W_OAm})}&{0}\\\hline
$\mathbf{W}^{l}_{A^v_1\to v_Q},\mathbf{W}^{l}_{A^m_1 \to m_Q}$&(\ref{W_Q})&0\\\hline
$\mathbf{W}^l_{\overline{h} \to Q}$&(\ref{W_Q})&$2d_cK$\\\hline
$\mathbf{W}^l_{\gamma},\mathbf{W}^l_{v\gamma \to \overline{h}}$&(\ref{W_h})&$d_cK$\\\hline
$\mathbf{W}_{\text{one}\to\overline{h}}$&(\ref{W_h})&0\\\hline
$\mathbf{W}^l_{m_{\overline{h}}\to\gamma},\mathbf{W}^l_{v_{\overline{h}}\to\gamma}$&(\ref{W_gamma})&$d_cK$\\\hline
$\mathbf{W}^l_{A_2^v \to v_\delta},\mathbf{W}^l_{A_2^m \to m_\delta}$&{(\ref{W_delta2z})}&{0}\\\hline
\!\!\!$\mathbf{W}^l_{y\!\to\! \delta},\mathbf{W}^l_{m_\delta \!\to\! m_\delta},\mathbf{W}^{l}_{v_\delta \!\to\! m_\delta},\mathbf{W}^l_{\lambda \!\to\! \delta}$\!\!\!&{(\ref{W_delta2z})}&{$L_tN$}\\\hline
$\mathbf{W}^l_{y\lambda\!\to\! z},\mathbf{W}^l_{mv\!\to\! z},\mathbf{W}^l_{\lambda\!\to\! z},\mathbf{W}_{{v_\delta}\!\to\! v_z}$&{(\ref{Wz})}&{$L_tN$}\\\hline
$\mathbf{W}^l_{m_z\to\lambda},\mathbf{W}^l_{y\to\lambda}$&{(\ref{W_lambda})}&{$2L_tN$}\\\hline
$\mathbf{W}^l_{v_z\to\lambda}$&{(\ref{W_lambda})}&{$L_tN$}\\
  \Xhline{1.2pt}
\end{tabular}
\label{comple}
\end{table}
\subsection{Computational Complexity}
It is emphasized that the weighting matrices are trained off-line, so that we can neglect the training complexity in the training period. Instead, we only focus on the online computational complexity of the DNN-MP-BSBL algorithm in the testing period. In addition, if all the non-zero entries in the weighting matrices are fixed as 1, the DNN-MP-BSBL algorithm is identical to the MP-BSBL algorithm, whose complexity  was thoroughly investigated in \cite{MPBSBL}. Therefore, we first present the computational complexity of the MP-BSBL algorithm as a benchmark.

In each iteration of the MP-BSBL algorithm, a total number of $(10L_t\!+\!10)Kd_c\!+\!(7d_r\!+\!14)L_tN\!+\!K\!+\!2$ multiplications are required. Specifically, $(10L_t+8)Kd_c$ multiplications are required for the calculation of $v^{l}_{\overline{h}_{k,d}}$ and $m^{l}_{\overline{h}_{k,d}}$ in (\ref{vmhkd}), $(7d_r+4)L_tN$ multiplications for $v^{l}_{\delta_n \to z_n}$ and $m^{l}_{\delta_n \to z_n}$ in (\ref{delta2z}), $(2d_c+1)K$ multiplications for $\hat{\gamma}^{l}_k$ in (\ref{gamma}), $8L_tN$ multiplications for $v^{l}_{z_n}$ and $m^{l}_{z_n}$ in (\ref{zn}), and $2L_tN+2$ multiplications for $\hat{\lambda}^{l}$ in (\ref{lambda}).

Compared with the MP-BSBL algorithm, the DNN-MP-BSBL algorithm incurs extra computational complexity when multiplying the non-zero weights in the weighting matrices. Then we analyze the number of extra multiplications, denoted by $\Delta_{\times}$, that is caused by each weighting matrix. Related results are listed in Table. \ref{comple}. Note that in Table. \ref{comple}, some weighting matrices will not lead to extra multiplications. For example, the weighting matrix $\mathbf{W}^{l}_{A^v_1\to v_Q}$ imposed on $\mathbf{O}^l_{A^v_1}$ in (\ref{W_Q}) is equivalently imposed on $\mathbf{P}^2$ in (\ref{W_OAv}). Therefore, we can replace $\mathbf{P}^2$ with $\mathbf{P}^2\times\mathbf{W}^{l}_{A^v_1\to v_Q}$ in (\ref{W_OAv}), and remove $\mathbf{W}^{l}_{A^v_1\to v_Q}$ from (\ref{W_Q}). In this way, $\Delta_{\times}=0$ for $\mathbf{W}^{l}_{A^v_1\to v_Q}$, and we can have the same results for $\mathbf{W}^{l}_{y\to A^m_1}$ and $\mathbf{W}^{l}_{{m_\delta}\to A^m_1}$ in (\ref{W_OAm}) and $\mathbf{W}^{l}_{A^m_1 \to m_Q}$ in (\ref{W_Q}), as well as $\mathbf{W}^l_{A_2^v \to v_\delta}$ and $\mathbf{W}^l_{A_2^m \to m_\delta}$ in (\ref{W_delta2z}). In addition, we have $\Delta_{\times}=0$ for $\mathbf{W}_{\text{one}\to\overline{h}}$ in (\ref{W_h}) since it serves as an additive bias term in the DNN, instead of a multiplicative weighting term. Furthermore, the weighting matrices in the first row of Table. \ref{comple} are imposed on the auxiliary layer $A^l_1$, which is far from the output of each iteration block. Therefore, the weights in these weighting matrices have negligible impacts on the final output, and we can fix these non-zero weights as 1 to reduce the computational complexity of the DNN-MP-BSBL algorithm.

According to the analysis, the DNN-MP-BSBL algorithm incurs $6Kd_c+13L_tN$ extra multiplications compared with the MP-BSBL algorithm. That is, a total of $(10L_t+16)Kd_c+(7d_r+27)L_tN+K+2$ multiplications are required by the DNN-MP-BSBL algorithm in each iteration, which is only slightly larger than that of the low-complexity MP-BSBL algorithm. For example, according to the configurations in Table. \ref{SimulationParameters}, the DNN-MP-BSBL algorithm  increases the computational complexity by only $4.3\%$, compared with the MP-BSBL algorithm. In addition, we also present the computational complexity of the BOMP estimator \cite{MPBSBL} and the LS-AMP-BSL estimator \cite{LSAMPSBL}, which are investigated for comparison in Section \ref{compar}. The overall number of multiplications in the BOMP estimator is $K^+L_td_cK+\frac{(K^++1)K^+}{2}L_td_c(N+1)+\frac{[(K^++1)K^+]^2d_c^3}{4}$, while that for the LS-AMP-BSL estimator is $N_{it}((9L_t+11)Kd_c+(7d_r+4)L_tN)+K^+L_td_c(N+1)+(K^+d_c)^3$, where $K^+$ is the number of active users. It is shown that the complexity of the BOMP estimator and the LS-AMP-BSL estimator increases with the square or cube of $K^+$. By contrast, the complexity of the DNN-MP-BSBL algorithm only increases linearly with $K$.
\section{Simulations}\label{simulationSec}
For the simulations, we consider a crowded NORA system with low-latency requirement, which is a common scenario for ultra-reliable low-latency communication (URLLC) and massive MTC (mMTC) applications in 5G and IoT. The congestion degree of the NORA system is characterized by the number of active users on each sub-carrier, which is proportional to the ratio $\frac{K}{N}$. Without loss of any generality, we assume a small number of available sub-carriers, i.e., $N=8$ to simulate the crowded RA scenarios with limited uplink resources. Other parameters related to the simulations and the DNN are listed in Table \ref{SimulationParameters}.

In order to satisfy the low-latency requirement, we investigate the UAD and CE accuracy of the DNN-MP-BSBL algorithm with a small number of iterations, i.e., $N_{it}\leq20$. For comparison, the NMSE performances of the LS-AMP-SBL esimator \cite{LSAMPSBL}, the BOMP estimator (with known active user number) \cite{MPBSBL} and the GA-MMSE estimator (with known user activity) are also provided. The NMSE performance of GA-MMSE estimator serves as the lower bound. It is shown that the proposed DNN-MP-BSBL algorithm could guarantee lower NMSE for CE and better UAD accuracy within a feasible number of iterations.
\begin{table}
\renewcommand\arraystretch{1.0}
\caption{Related Parameters for Simulations}\vspace{-0.1cm}
\centering
\begin{tabular}{ccc}
  \Xhline{1.2pt}
Parameter&Symbol&Value\\
  \hline
User number&$K$&110\\
Subcarrier number&$N$&8\\
Pilot length&$L_t$&11\\
Spreading factor&$d_c$&4\\
Activation probability for each user&$P_a$&0.1\\
UAD threshold&$\gamma_{th}$&0.1\\
Size of training set&&$10^5$\\
Size of test set&&$10^5$\\
Size of mini-batch&&200\\
Epoch number&&20\\
Learning rate&&$10^{-3}$\\
  \Xhline{1.2pt}
\end{tabular}
\label{SimulationParameters}
\end{table}
\subsection{Convergence}
\begin{figure}
  \centering
  \includegraphics[width=0.9\linewidth]{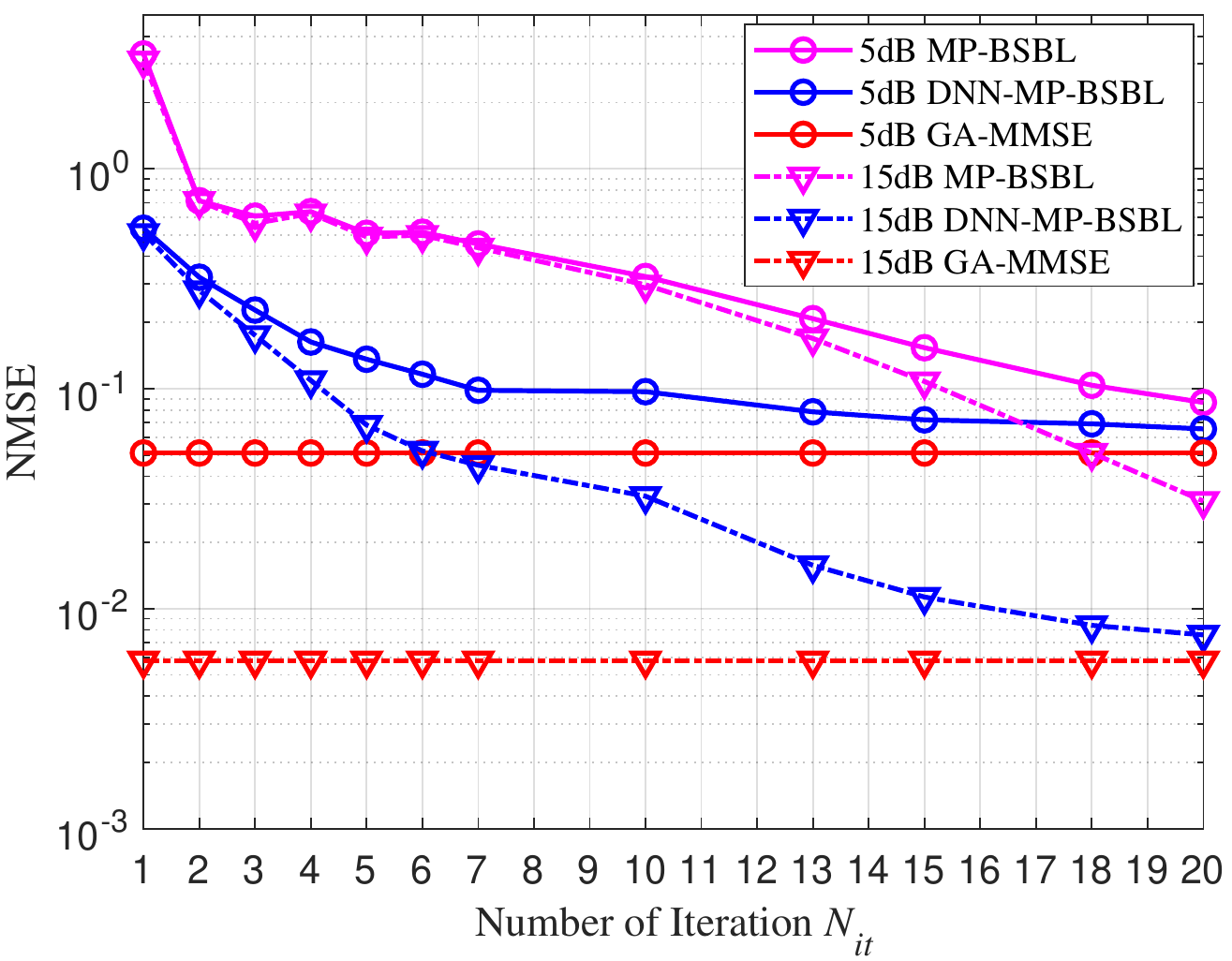}\vspace{-0.2cm}
  \caption{Convergence with iteration for the proposed DNN-MP-BSBL algorithm and the MP-BSBL algorithm \cite{MPBSBL}.}\label{Convergence}
\end{figure}
The convergence performances of the proposed DNN-MP-BSBL algorithm and the MP-BSBL algorithm \cite{MPBSBL} are illustrated in Fig. \ref{Convergence}, with different number of iterations and different SNR. Also included is the NMSE of the GA-MMSE estimator. It is shown in Fig. \ref{Convergence} that for a small number of iterations, the MP-BSBL algorithm converges slowly and exhibits limited improvement for higher SNR. By contrast, the proposed DNN-MP-BSBL algorithm exhibits better NMSE performance and faster convergence. For different SNR, the DNN-MP-BSBL algorithm with 10 iterations could achieve the same NMSE as the MP-BSBL algorithm with 20 iterations. With 20 iterations, the DNN-MP-BSBL algorithm could closely approach the lower bound, and the gaps to the bound are $1.3\times 10^{-2}$ and $1.7\times 10^{-3}$ for SNR$=$5dB and 15dB, respectively. In addition, the improvement by the DNN-MP-BSBL algorithm is more prominent with a higher SNR. For example, the DNN-MP-BSBL algorithm could improve the NMSE performance of the MP-BSBL algorithm by almost one magnitude order when SNR$=$15dB.
\subsection{CE and UAD Accuracy With Different $P_a$}
\begin{figure}
  \centering
  \includegraphics[width=0.9\linewidth]{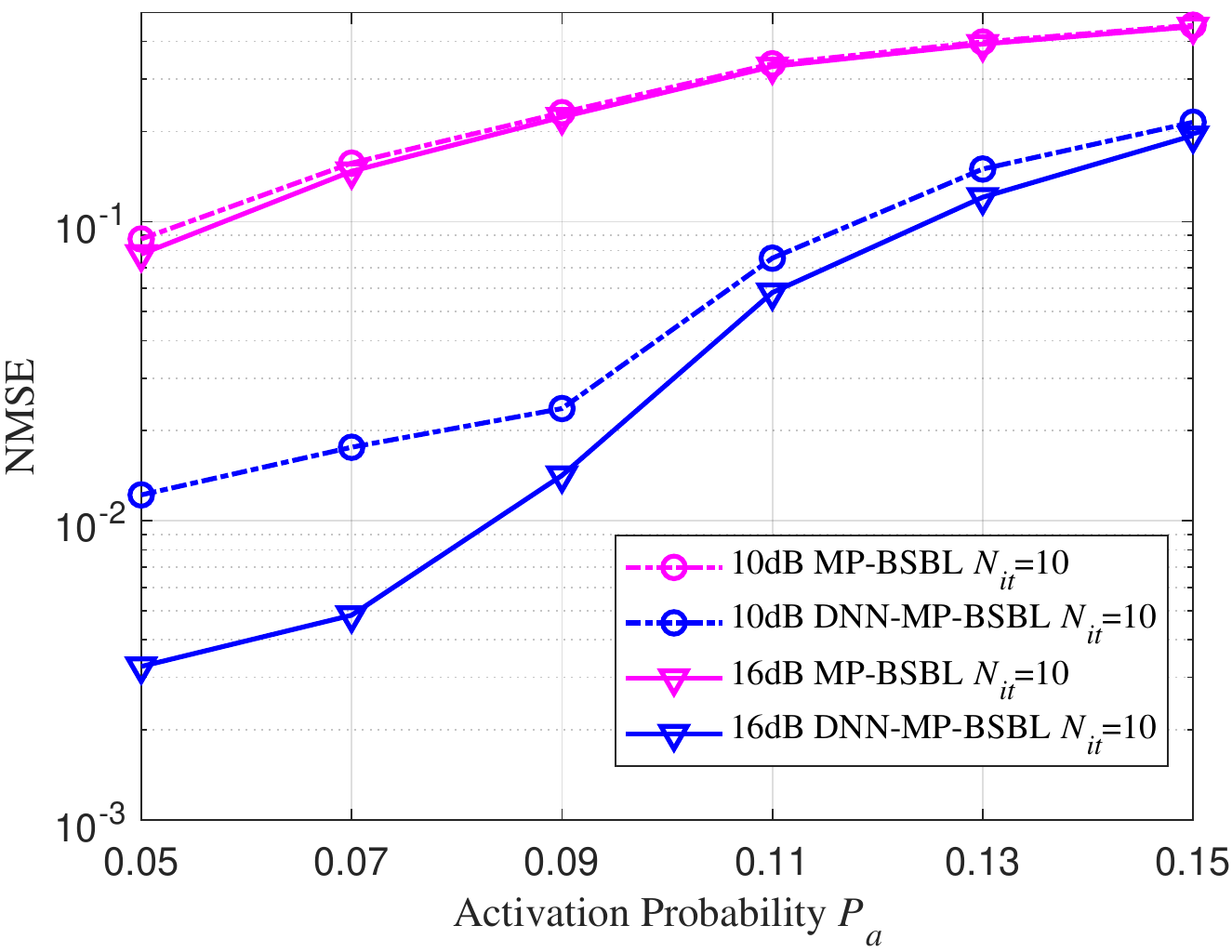}\vspace{-0.2cm}
  \caption{NMSE performance for the proposed DNN-MP-BSBL algorithm and the MP-BSBL algorithm \cite{MPBSBL} with different activation probability $P_a$.}\label{changepanmse}
\end{figure}
\begin{figure}
  \centering
  \includegraphics[width=0.9\linewidth]{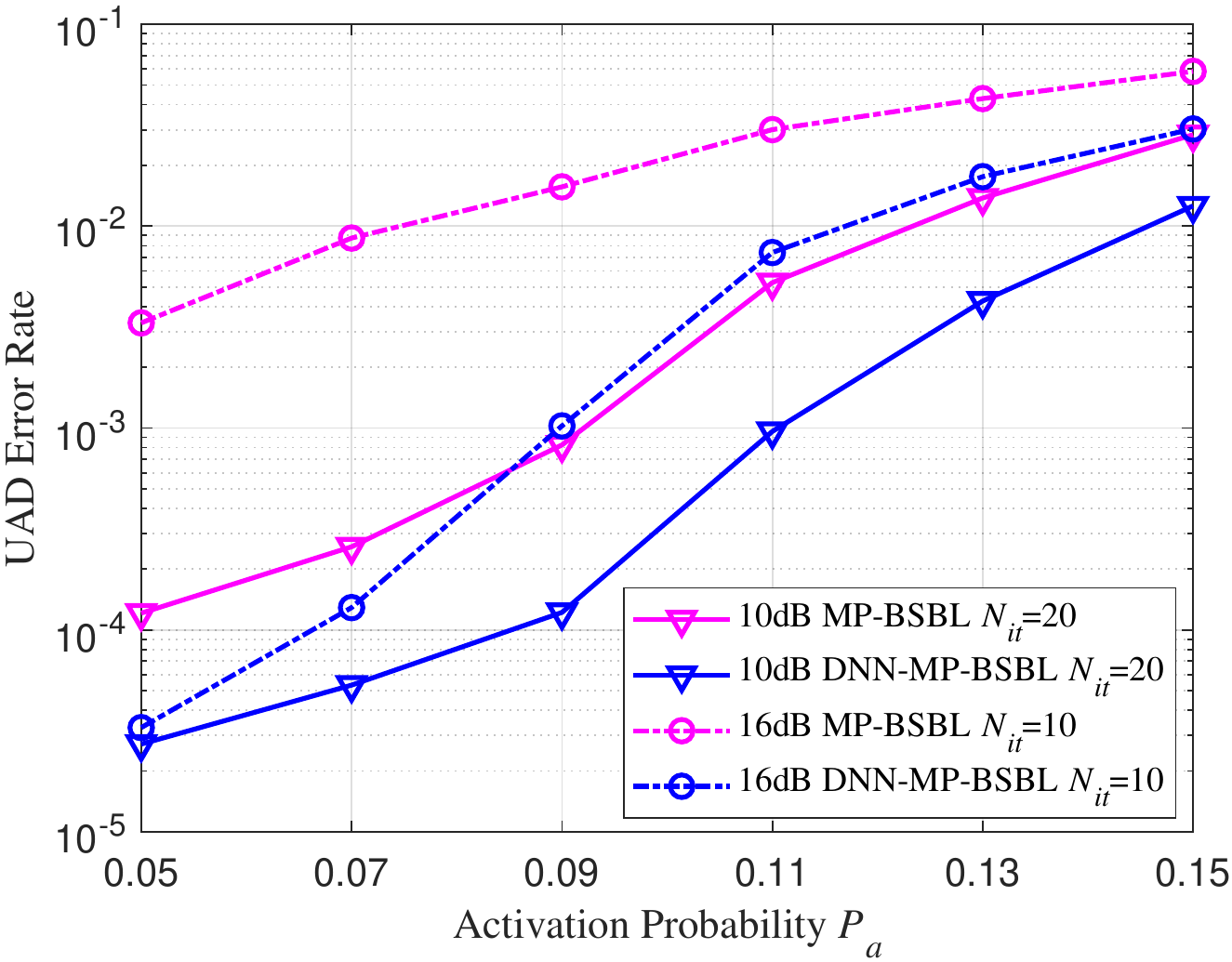}\vspace{-0.2cm}
  \caption{UAD accuracy for the proposed DNN-MP-BSBL algorithm and the MP-BSBL algorithm \cite{MPBSBL} with different activation probability $P_a$.}\label{changepape}
\end{figure}
The NMSE performance of the proposed DNN-MP-BSBL algorithm is investigated with different activation probability $P_a$, and the simulation results are illustrated in Fig. \ref{changepanmse}. It is shown that when $P_a$ increases, the NORA system is crowded with more active users. As a result, the NMSE performance deteriorates for both the DNN-MP-BSBL algorithm and the MP-BSBL algorithm. In addition, the NMSE performance of the MP-BSBL algorithm shows no improvement when SNR increases from 10dB to 16dB. By contrast, the proposed DNN-MP-BSBL algorithm could always guarantee improvement on the NMSE performance with different SNR and different $P_a$. Particularly, the improvement is more prominent with lower $P_a$ and higher SNR. For example, the DNN-MP-BSBL algorithm could improve the NMSE by almost one magnitude order when SNR=10dB and $P_a$ is relatively low. Furthermore, the NMSE can be further lowered by the DNN-MP-BSBL algorithm when SNR increases to 16dB.

In the DNN-MP-BSBL algorithm, weights are also imposed on the MF message update in (\ref{W_gamma}) to alleviate the influence of correlated Gaussian messages and improve the UAD accuracy. We also investigate the UAD accuracy of the DNN-MP-BSBL algorithm with different $P_a$, and the results are shown in Fig. \ref{changepape}. It is shown that the proposed DNN-MP-BSBL algorithm could improve the UAD accuracy of the MP-BSBL algorithm for different $P_a$, SNR, and different number of iterations. Similarly, this improvement on UAD accuracy is more prominent when $P_a$ is relatively low and SNR is high. For example, the DNN-MP-BSBL algorithm could improve the UAD error rate by two magnitude orders when SNR=16dB and $P_a\leq0.1$.
\subsection{CE Accuracy With Different SNR}\label{compar}
\begin{figure}
  \centering
  \includegraphics[width=0.9\linewidth]{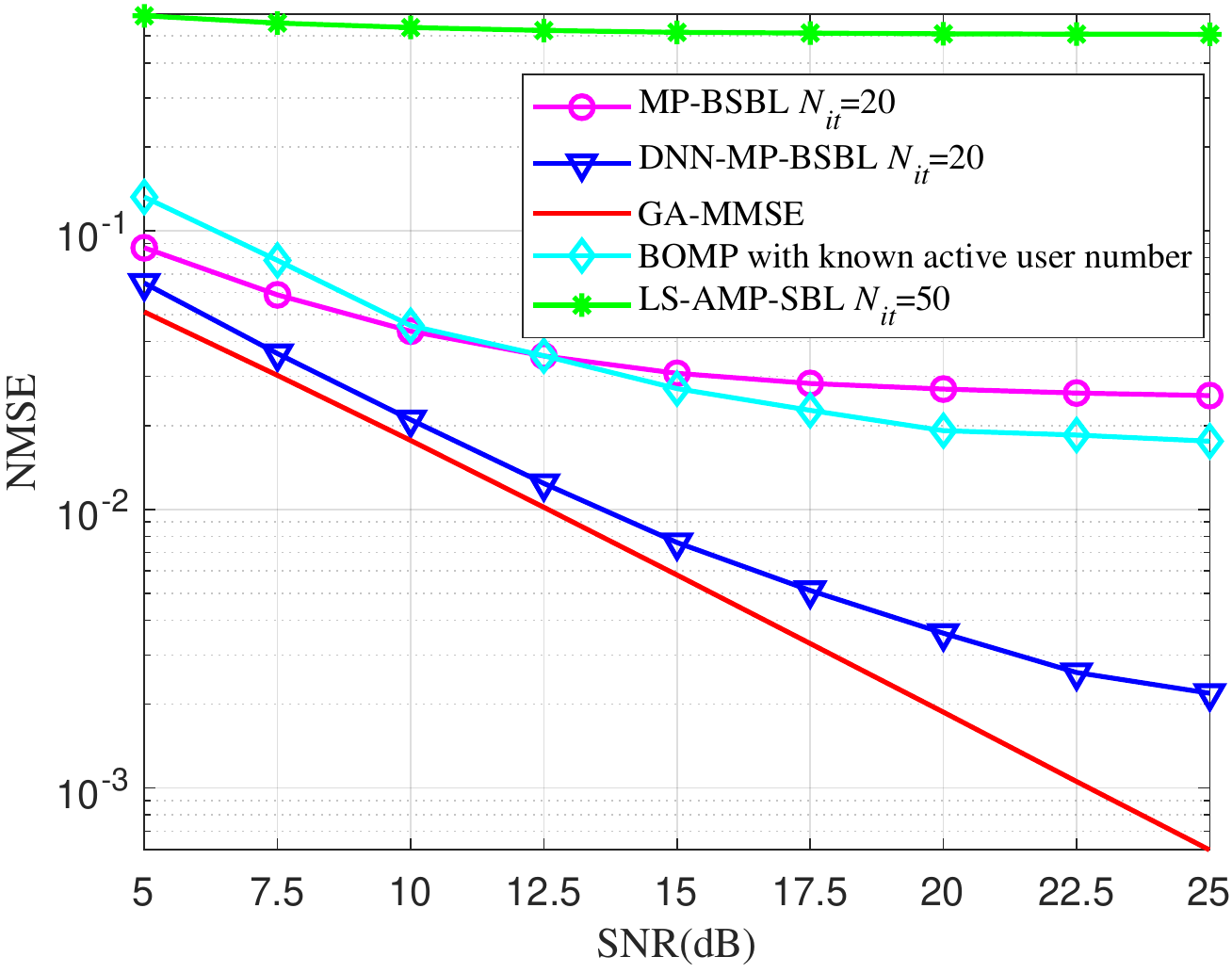}\vspace{-0.3cm}
  \caption{NMSE performance for CE with different SNR.}\label{changeSNR}
\end{figure}
Finally, we compare the proposed DNN-MP-BSBL algorithm with the LS-AMP-BSL estimator \cite{LSAMPSBL}, the BOMP estimator (with known active user number) \cite{MPBSBL}, and the GA-MMSE estimator (with known user activity). The simulation results are illustrated in Fig. \ref{changeSNR} and the NMSE performance of the GA-MMSE estimator serves as the lower bound.

As shown in Fig. \ref{changeSNR}, in crowded NORA systems, both the MP-BSBL algorithm and the BOMP estimator diverge from the NMSE lower bound as SNR increases, and the LS-AMP-SBL algorithm fails to work even with 50 iterations. By contrast, the DNN-MP-BSBL algorithm could closely approach the lower bound within a wide range of SNR. Therefore, the DNN-MP-BSBL algorithm requires fewer iterations and provides better NMSE performance, indicating its advantages in crowded NORA system with low-latency requirement.
\section{Conclusions}\label{conclusions}
A DNN-MP-BSBL algorithm was proposed in this paper for the joint UAD and CE problem in grant-free NORA systems. The iterative message passing process is transferred from a factor graph to a DNN, while weights are imposed on the messages and trained to improve the UAD and CE accuracy. It was verified that the weighted message passing in the DNN-MP-BSBL algorithm could alleviate the convergence problem of the MP-BSBL algorithm in crowded NORA systems. Simulation results showed that the NMSE performance of the DNN-MP-BSBL algorithm could approach the lower bound in a feasible number of iterations, indicating its advantages for low-latency NORA systems.
\bibliographystyle{IEEEtran}


\begin{IEEEbiography}[{\includegraphics[width=1in,height=1.25in,clip,keepaspectratio]{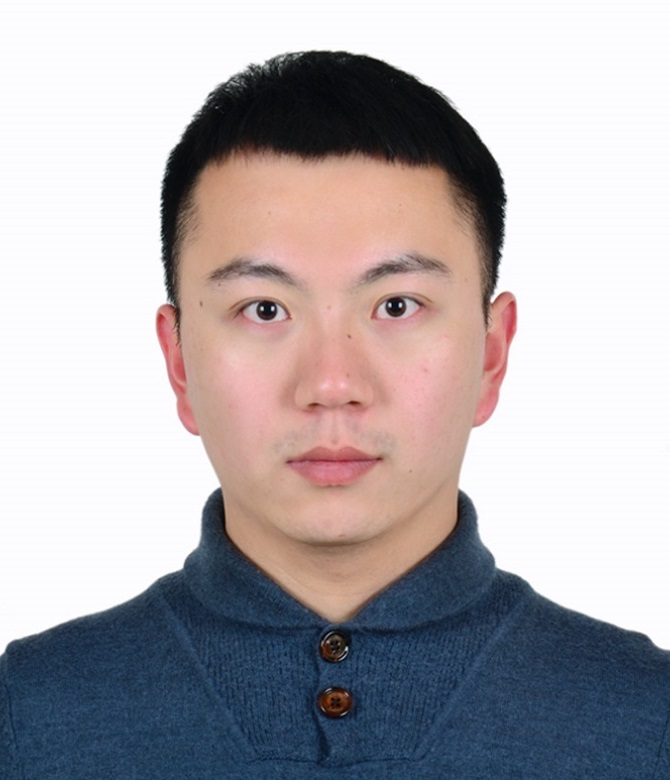}}]{Zhaoji Zhang}  received the B.Eng. degree in telecommunication engineering from Ocean University of China in 2015. He is currently pursuing the Ph.D. degree in communication and information systems with the State Key Laboratory of Integrated Services Networks, Xidian University, Xi'an, China. From 2018 to 2019, he was with the School of Electrical and Electronic Engineering, Nanyang Technological University, Singapore, as a visiting PhD student. His research interests include channel coding, message passing algorithms, deep neural network, and random access in M2M communications.
\end{IEEEbiography}

\begin{IEEEbiography}[{\includegraphics[width=1in,height=1.25in,clip,keepaspectratio]{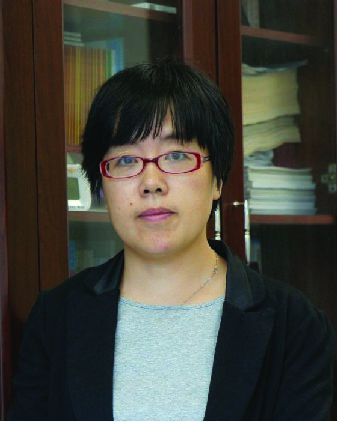}}]{Ying Li} received the B.S. degree in telecommunication engineering and the Ph.D. degree in communication and information systems from Xidian University, Xi'an, China, in 1995 and 2005, respectively. From 2011 to 2012, she was with the University of California at Davis, Davis, CA, USA, as a Visiting Scholar. She is currently a Professor with Xidian University. Her current research interests are on design and analysis for wireless systems, including channel coding, wireless network communications, interference processing, and MIMO techniques.
\end{IEEEbiography}

\begin{IEEEbiography}[{\includegraphics[width=1in,height=1.25in,clip,keepaspectratio]{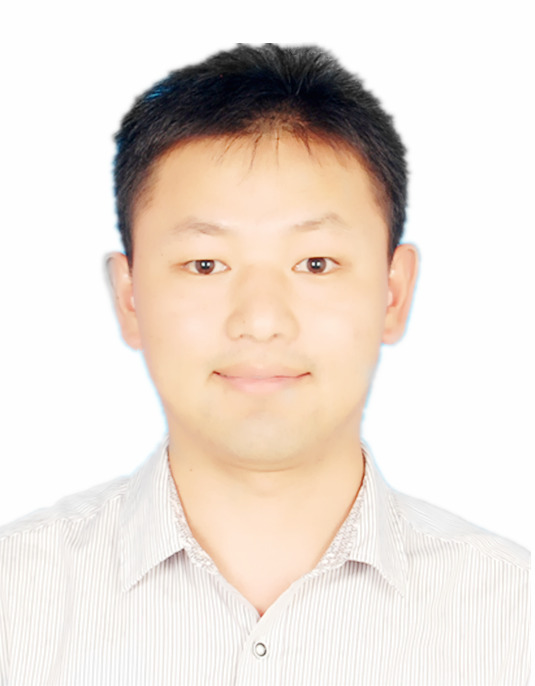}}]{Chongwen Huang} received the B.Sc. degree from Nankai University in 2010, and the M.Sc. degree from the University of Electronic Science and Technology of China, Chengdu, in 2013. In 2015, he started his Ph.D. journey at Singapore University of Technology and Design, Singapore, under the supervision of Prof. C. Yuen and Prof. M. Debbah. He joined the Institute of Electronics, Chinese Academy of Sciences, IECAS, Beijing, as a Research Engineer in 2013. His main research interests are focused on deep learning, machine learning for wireless communication and signal processing, large intelligent surface for 5G and beyond, millimeter wave communications, multi-user information theory, and compressed sensing.
\end{IEEEbiography}

\begin{IEEEbiography}[{\includegraphics[width=1in,height=1.25in,clip,keepaspectratio]{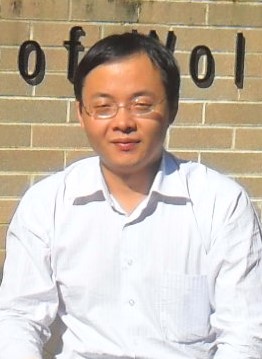}}]{Qinghua Guo (S'07-M'08-SM'18)}  received the B.E. degree in electronic engineering and the M.E. degree in signal and information processing from Xidian University, Xi'an, China, in 2001 and 2004, respectively, and the Ph.D. degree in electronic engineering from the City University of Hong Kong, Kowloon, Hong Kong, in 2008. He is currently an Associate Professor with the School of Electrical, Computer and Telecommunications Engineering, University of Wollongong, Wollongong, NSW, Australia, and an Adjunct Associate Professor with the School of Engineering, The University of Western Australia, Perth, WA, Australia. His research interests include signal processing and telecommunications. He was a recipient of the Australian Research Council’s inaugural Discovery Early Career Researcher Award in 2012.
\end{IEEEbiography}

\begin{IEEEbiography}[{\includegraphics[width=1in,height=1.25in,clip,keepaspectratio]{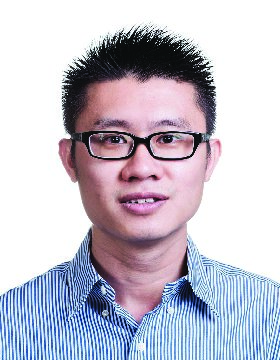}}]{Chau Yuen} received the B.Eng. and Ph.D. degrees from Nanyang Technological University, Singapore, in 2000 and 2004, respectively. He was a PostDoctoral Fellow at Lucent Technologies Bell Labs, Murray Hill, NJ, USA, in 2005. He was a Visiting Assistant Professor at The Hong Kong Polytechnic University in 2008. From 2006 to 2010, he was a Senior Research Engineer at the Institute for Infocomm Research (I2R, Singapore), where he was involved in an industrial project on developing an 802.11n Wireless LAN system, and participated actively in 3Gpp Long Term Evolution (LTE) and LTE-Advanced (LTE-A) Standardization. He has been with the Singapore University of Technology and Design as an Assistant Professor since 2010. He is a recipient of the Lee Kuan Yew Gold Medal, the Institution of Electrical Engineers Book Prize, the Institute of Engineering of Singapore Gold Medal, the Merck Sharp \& Dohme Gold Medal, and twice the recipient of the Hewlett Packard Prize. He received the IEEE Asia-Pacific Outstanding Young Researcher Award in 2012. He serves as an Editor for the IEEE TRANSACTION ON COMMUNICATIONS and the IEEE TRANSACTIONS ON VEHICULAR TECHNOLOGY and was awarded the Top Associate Editor from 2009 to 2015.
\end{IEEEbiography}

\begin{IEEEbiography}[{\includegraphics[width=1in,height=1.25in,clip,keepaspectratio]{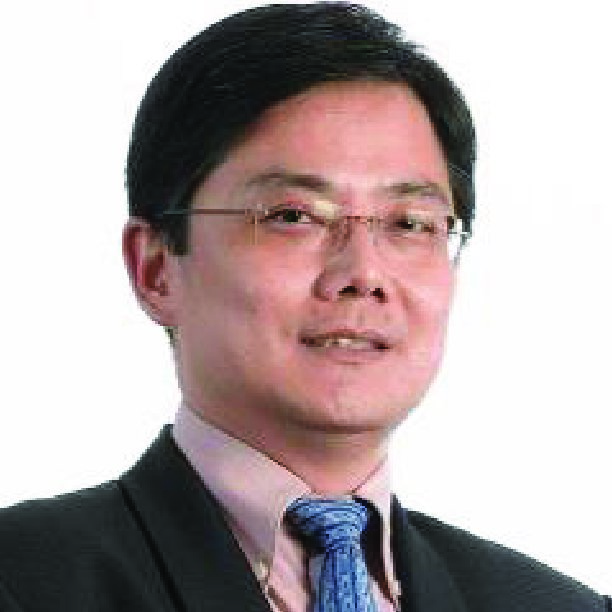}}]{Yong Liang Guan} received the Ph.D. degree from the Imperial College of London, U.K., and the B.Eng. degree (Hons.) from the National University of Singapore. He is currently a tenured Associate Professor with the School of Electrical and Electronic Engineering, Nanyang Technological University, Singapore. His research interests broadly include modulation, coding and signal processing for communication systems, and data storage systems. He is an Associate Editor of the IEEE TRANSACTIONS ON VEHICULAR TECHNOLOGY. His homepage is http://www3.ntu.edu.sg/home/eylguan.
\end{IEEEbiography}

\end{document}